\documentclass[
    journal=cmatex,
    manuscript=article,
    layout=twocolumn,
]{achemso}

\usepackage[version=4]{mhchem}
\usepackage{amsmath}
\usepackage{amssymb}
\usepackage{glossaries}
\usepackage{graphicx}
\usepackage{tabularx}
\usepackage{siunitx}
\usepackage{hyperref}

\newcommand \Title{
    Octahedral tilt-driven phase transitions in \texorpdfstring{\ce{BaZrS3}}{BaZrS3} chalcogenide perovskite
}
\hypersetup{
    pdfauthor={P. Kayastha, E. Fransson, P. Erhart, L. Whalley},
    pdftitle={\Title},
    pdffitwindow=false,
    pdfstartview={FitH},
    pdfnewwindow=true,
    colorlinks=true,
    linkcolor=black,
    citecolor=black,
    filecolor=black,
    urlcolor=black,
}

\graphicspath{{./figures/}}
\SectionNumbersOn

\setacronymstyle{long-short}
\newacronym{dft}{DFT}{density functional theory}
\newacronym{rms}{RMS}{root mean square}
\newacronym{gpu}{GPU}{graphics processing unit}
\newacronym{md}{MD}{molecular dynamics}
\newacronym{mlp}{MLP}{machine-learned potential}
\newacronym{nep}{NEP}{neuroevolution potential}
\newacronym{sm}{SM}{Supplemental Material}
\newacronym{soap}{SOAP}{smooth overlap of atomic positions}
\newacronym{xc}{XC}{exchange-correlation}
\newacronym{xrd}{XRD}{X-ray diffraction}

\newacronym{scan}{SCAN}{strongly constrained and appropriately normed}
\newacronym{wham}{WHAM}{weighted histogram analysis method}
\newacronym{ti}{TI}{thermodynamic integration}

\newcommand{\ase}{\textsc{ase}}
\newcommand{\calorine}{\textsc{calorine}}

\newcommand{\hmn}[1]{
  \ensuremath{\begingroup\setupHMN #1\endgroup}%
}

\newcommand{\setupHMN}{%
  \doHMN{-}{\HMNoverline}%
  \doHMN{*}{\HMNminverse}%
  \doHMN{i}{\infty}
}

\newcommand{\doHMN}[2]{%
  \begingroup\lccode`~=`#1
  \lowercase{\endgroup\let~}#2%
  \mathcode`#1="8000
}

\newcommand{\HMNminverse}[1]{\frac{#1}{m}}
\newcommand{\HMNoverline}[1]{\mkern1mu\overline{\mkern-1mu#1\mkern-1mu}\mkern1mu}

\DeclareSIUnit\angstrom{\text{Å}}
\DeclareSIUnit\atom{\text{atom}}

\newcommand{\addchalmers}{Department of Physics, Chalmers University of Technology, SE-41296, Gothenburg, Sweden}

\newcommand{\addnorthumbria}{Department of Mathematics, Physics and Electrical Engineering, Northumbria University, Newcastle upon Tyne, NE1 8QH, United Kingdom}

\title{\Title}

\author{Prakriti Kayastha}
\affiliation{\addnorthumbria}
\author{Erik Fransson}
\affiliation{\addchalmers}
\author{Paul Erhart}
\affiliation{\addchalmers}
\author{Lucy Whalley}
\affiliation{\addnorthumbria}
\email{l.whalley@northumbria.ac.uk}

\begin{document}

\begin{tocentry}
\includegraphics[width=8.25 cm]{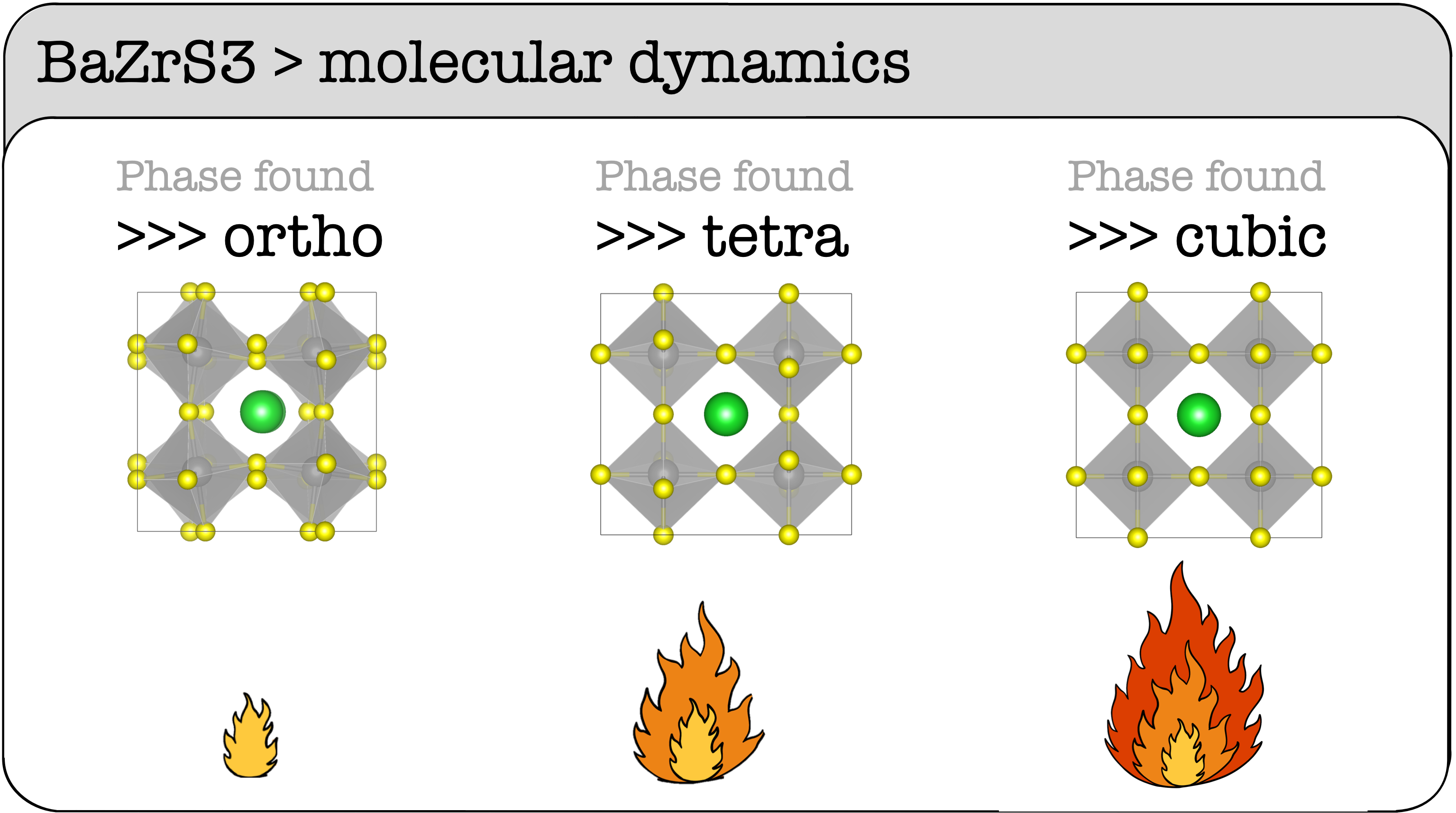}
\end{tocentry}

\begin{abstract}
Chalcogenide perovskites are lead-free materials for potential photovoltaic or thermoelectric applications.
\ce{BaZrS3} is the most studied member of this family due to its superior thermal and chemical stability, desirable optoelectronic properties, and low thermal conductivity. 
Phase transitions of \ce{BaZrS3} remain underexplored in the literature, as most experimental characterizations of this material have been performed at ambient conditions where the orthorhombic \hmn{Pnma} phase is reported to be stable. 
In this work, we study the dynamics of \ce{BaZrS3} across a range of temperatures and pressures using an accurate machine-learning interatomic potential trained with data from hybrid density functional theory calculations. 
At \qty{0}{\pascal}, we find a first-order phase transition from the orthorhombic to tetragonal \hmn{I4/mcm} phase at \qty{610}{\kelvin}, and a second-order transition from the tetragonal to the cubic \hmn{Pm-3m} phase at \qty{880}{\kelvin}. 
The tetragonal phase is stable over a larger temperature range at higher pressures. 
To confirm the validity of our model we compare our results with a range of published experimental data and report a prediction for the X-ray diffraction pattern as a function of temperature. 
\end{abstract}

\maketitle
\newpage

Chalcogenide perovskites have gained relevance as lead-free photovoltaic absorber materials as they exhibit strong light absorption and dielectric screening alongside desirable defect properties \cite{sun2015chalcogenide, sopiha2022chalcogenide, tiwari2021chalcogenide, jaramillo2019praise,choi2022emerging,nishigaki2020extraordinary,ravi2021colloidal,wu2021defect, yuan2024assessing}.
\ce{BaZrS3} is the most studied member of this family with research efforts including material synthesis at moderate temperatures, band-gap engineering, and a proof-of-concept solar cell \cite{yu2021chalcogenide, yang2022low, pradhan2023synthesis, comparotto2022synthesis, turnley2022solution,sharma2021bandgap, sadeghi2023expanding, meng2016alloying,dallas2024exploring, Agarwal2025from}.
\ce{BaZrS3} has also been explored as a potential thermoelectric material as it displays fast electronic transport coupled with low thermal conductivity, leading to record-high $zT$ values among reported halide and chalcogenide perovskite materials \cite{osei2021examining,osei2019understanding,yang2024chalcogenide}.

Whilst \ce{BaZrS3} is reported to retain stability in a perovskite structure above \qty{1000}{\kelvin}, the vast majority of materials characterization is carried out at ambient conditions \cite{jaykhedkar2023temperature,comparotto2020chalcogenide, xu2022enhancing, yu2021chalcogenide,yang2022low}.
At room temperature, the consensus is that \ce{BaZrS3} is stable in an orthorhombic \hmn{Pnma} perovskite structure, as confirmed by both experimental and computational studies \cite{niu2019crystal, comparotto2022synthesis, mukherjee2023interplay,gross2017stability,
filippone2020discovery, osei2019understanding, kayastha2023high}.
Above this temperature, the picture is less clear.
Recent temperature-dependent \gls{xrd} measurements show a discontinuous change in the lattice parameters at high temperature, indicating a first-order phase transition \cite{jaykhedkar2023temperature, bystricky2024thermal,Jaiswal2024High}. A temperature-dependent Raman spectroscopy study does not confirm this observation \cite{ye2024differing}, likely due to the structural and dynamic similarity of perovskite phases coupled with significant thermal broadening. 

Many \ce{ABX3} perovskites undergo tilt-driven phase transitions to form lower-symmetry polymorphs with antiferrodistortive displacement patterns \cite{redfern1996high}.
The prototypical high-temperature perovskite phase is a cubic structure.
As the temperature is reduced, lower-symmetry tetragonal and orthorhombic perovskite phases can be formed through tilting of the \ce{BX6} octahedra \cite{fransson2023phase, fransson2023revealing}.
Therefore, it is likely that there are transitions to higher symmetry phases at temperatures above ambient for \ce{BaZrS3}.
Phase transitions may occur before reaching the elevated temperatures required for \ce{BaZrS3} synthesis ($>$\qty{850}{\kelvin}) \cite{comparotto2022synthesis}.
If this is the case it follows that samples grown at high temperature may include mixtures of polymorphs, as has been observed for halide perovskites.\cite{Dubajic2024nano,Weadock2023nature}
A phase transition within the operating temperature range for thermoelectric generators (\qtyrange{400}{1100}{\kelvin}) is also possible.
An understanding of the exact \ce{BaZrS3} perovskite structure is important as even small changes to structure can impact key functional properties including the band gap \cite{Glazer1975simple,Linaburg2017}.
Anharmonic dynamics are also crucial for quantitative predictions of electron-phonon coupling and related optical properties.

In this work, we use \gls{md} to sample the anharmonic free energy surface and simulate the finite-temperature dynamics of the \ce{BaZrS3} perovskite. 
We accelerate the calculation of free energies, atomic forces, and stress tensors necessary for \gls{md} by constructing a machine-learning interatomic potential using reference data from \gls{dft} calculations using a hybrid exchange-correlation functional.
We use known group-subgroup relationships to systematically identify which octahedral tilt patterns can be accessed during phase transitions.
We identify two phase transitions in \ce{BaZrS3} at \SI{0}{\pascal}: a first-order orthorhombic \hmn{Pnma} to tetragonal \hmn{I4/mcm} transition at \qty{610}{\kelvin} and a second-order tetragonal to cubic \hmn{Pm-3m} phase transition at \qty{880}{\kelvin}.
We construct a phase diagram for \ce{BaZrS3} across a pressure and temperature range of \qtyrange{-4}{10}{\giga\pascal} and \qtyrange{0}{1200}{\kelvin}, respectively.
Lastly, we predict the Raman spectra and temperature-dependent X-ray diffraction pattern, and compare our predictions against published experimental data. 

\section*{Methods}

A machine-learning interatomic potential was constructed using the \gls{nep} method implemented in the \textsc{gpumd} package \cite{FanWanYin22}. The \ase{} and \calorine{} packages were used to prepare the training structures, set up \gls{md} simulations, and post-process the results.\cite{Larsen2017, calorine}
The training set consists of \num{1187} perovskite structures. This includes cubic, tetragonal, and orthorhombic phases with applied strain or small random displacements, all 15 Glazer-tilt structures,\cite{glazer1972classification,howard1998group} and snapshots from NPT \gls{md} simulations. 
The training set also contains \num{92} Ruddlesden-Popper structures, which will be the subject of a future publication.
Symmetry-constrained geometry relaxations as implemented in \ase{} were performed until the maximal force component was below \qty{e-3}{\electronvolt\per\angstrom} \cite{Larsen2017}.  
\Gls{dft} calculations were performed using the \textsc{fhi-\rm{aims}} code and the HSE06 exchange-correlation functional \cite{blum2009ab,krukau2006influence}. 
The root mean squared training errors were \qty{1.8}{\milli\electronvolt\per\atom}, \qty{72.2}{\milli\electronvolt\per\angstrom}, and \qty{28.9}{\milli\electronvolt\per\atom} for formation energies, atomic forces, and virials, respectively; see \autoref{sfig:loss_curves} for the loss curves and \autoref{sfig:parity_plots} for the parity plots. 
Harmonic phonon dispersions were evaluated using the \textsc{phonopy} package with a \numproduct{2x2x2} supercell and a \qty{0.01}{\angstrom} displacement \cite{togo2023first}. 
For a comparison of the \gls{nep}-calculated and \gls{dft}-calculated harmonic phonon dispersions see
\autoref{sfig:Pnma_DFT_vs_NEP_phonons}, \autoref{sfig:I4_mcm_DFT_vs_NEP_phonons} and \autoref{sfig:Pm3m_DFT_vs_NEP_phonons}. 

Heating and cooling simulations with supercells of \num{40960} atoms were run in the NPT ensemble in the temperature range of \qtyrange{0}{1200}{\kelvin} and a pressure range of \qtyrange{-4}{10}{\giga\pascal} using a timestep of \qty{1}{\femto\second}. 
To identify the symmetry group formed, atomic displacements were projected onto the octahedral-tilt phonon eigenvectors of the cubic structure, as outlined in Ref.~\citenum{FraRosEriRahTadErh23}. 
Mode projections amplitudes for the \gls{nep}-relaxed and \gls{dft}-relaxed structures are given in \autoref{tab:mode_projection_NEP} and \autoref{tab:mode_projection_DFT}, respectively.
Free energy calculations were carried out using \gls{ti} with an Einstein crystal as reference Hamiltonian \cite{FreLad84}. 
To calculate the X-ray diffraction pattern $I(\boldsymbol{\theta})$ the \textsc{dynasor} package was used to post-process NVT \gls{md} simulations.\cite{FraSlaErhWah2021}
For more computational details see the Supplementary Information.

\begin{figure}
    \centering
    \includegraphics[width=\linewidth]{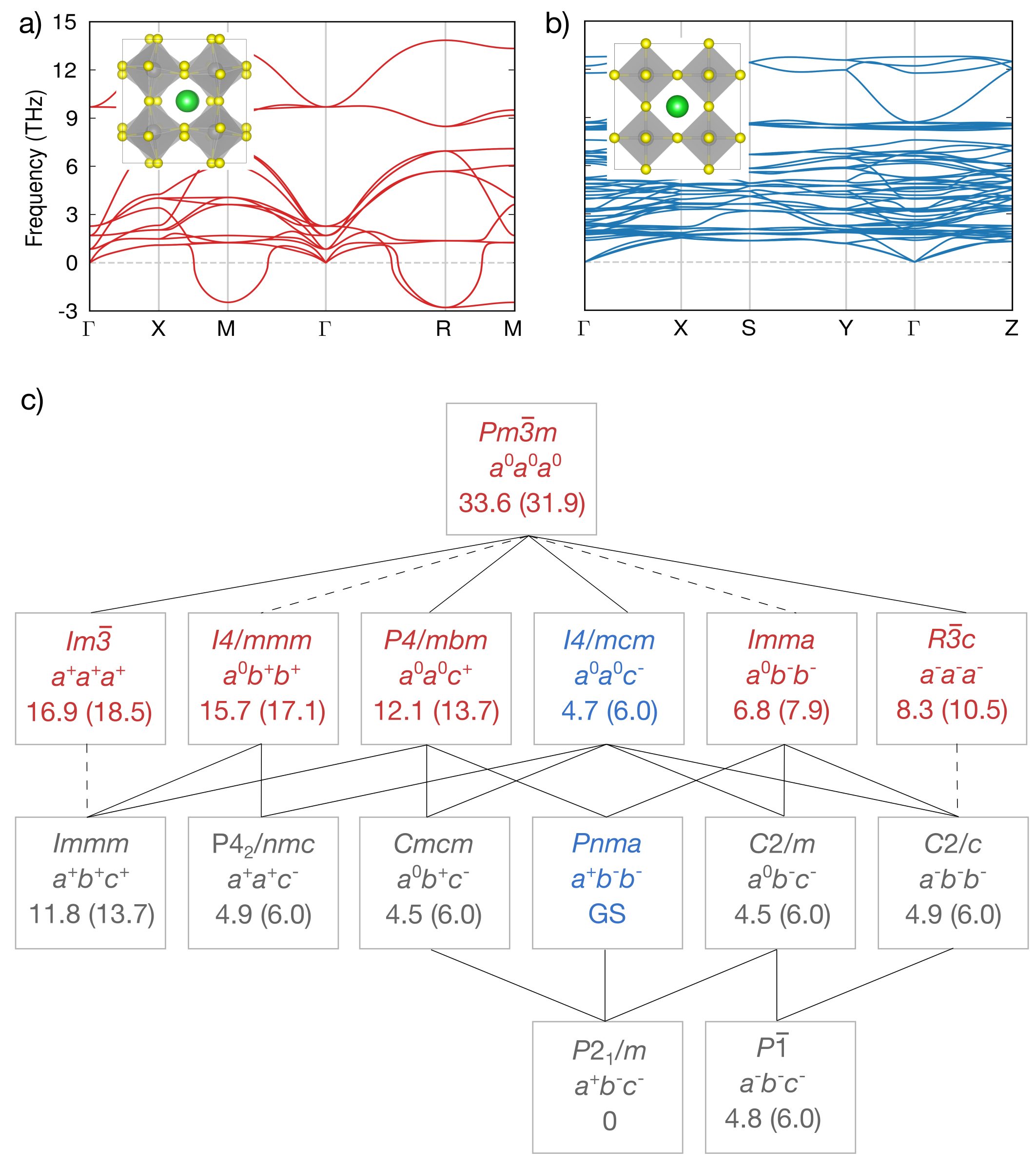}
    \caption{\Gls{dft}-calculated crystal and phonon band structures of the (a) orthorhombic \hmn{Pnma}
    (b) and cubic \hmn{Pm-3m} phases.
    Green, grey, and yellow spheres represent Ba, Zr, and S atoms respectively.
    () Group-subgroup relationships and formation energies.
    The space group, Glazer notation, and formation energy are specified for each phase accessible through octahedral tilting.
    The formation energies (in \unit{\milli\electronvolt\per\atom}) with respect to the ground state (GS) \hmn{Pnma} phase obtained from \gls{dft} and \gls{nep} model calculations (in parentheses) are reported in the bottom rows.
    Connecting lines indicate group-subgroup relationships.
    Dashed lines indicate transitions that must be first order according to Landau theory \cite{howard1998group}.
    Red text denotes the presence of phonon modes with an imaginary frequency, indicating dynamic instability at \qty{0}{\kelvin}.
    Blue text denotes dynamic stability.
    Grey text denotes that the phase is symmetrically equivalent to a supergroup structure after relaxation. There may still be a small discrepancy in formation energy, which is discussed in the Supplementary Information.
    Figure adapted from Howard and Stokes with permission from the International Union of Crystallography \cite{howard1998group}.
    } 
    \label{fig:Glazer_space}
\end{figure}

In \autoref{fig:Glazer_space}a and \autoref{fig:Glazer_space}b we plot the harmonic phonon dispersions and crystal structures of \ce{BaZrS3} in the cubic \hmn{Pm-3m} phase and experimentally observed \hmn{Pnma} phase. 
The aristotype cubic \hmn{Pm-3m} phase is the simplest perovskite form. 
However, perovskites often adopt lower-symmetry, distorted non-cubic phases \cite{tilley2016perovskites}. 
Distortions in the cubic perovskite give rise to a wide range of structures which can be classified into three categories: (i) \ce{BX6} octahedral tilting; (ii) distortions of the \ce{BX6} octahedra; and (iii) B-site cation displacements \cite{lufaso2004jahn, howard2005structures}.
Octahedral tilting leads to 15 possible space groups as identified by Glazer \cite{glazer1972classification}.

\ce{BaZrS3} in the cubic \hmn{Pm-3m} phase ($a^0 a^0 a^0$ in Glazer notation) is dynamically unstable indicating the presence of a lower-symmetry stable structure at \qty{0}{\kelvin}\cite{pallikara2022physical}.
The imaginary phonon modes at the M point of the Brillouin zone correspond to in-phase ($^+$) tilting of the \ce{ZrS6} octahedra and are described with the irrep M$_2^+$ (for a unit cell with an origin at the Ba-site) \cite{Glazer2021Journey}.
The imaginary modes at the R point correspond to out-of-phase ($^-$) tilting and have irrep R$_5^-$. 
Both modes are triply degenerate.
Distortions along one M-mode and two perpendicular R-modes result in a dynamically stable orthorhombic \hmn{Pnma} phase ($a^+ b^- b^-$). 
The dynamic and energetic stability of the orthorhombic phase at \qty{0}{\kelvin} is in agreement with previous experimental and \gls{dft} studies reporting it to be stable at low temperatures \cite{lelieveld1980sulphides, okai1988preparation, perera2016chalcogenide}.

Distortions along linear combinations of the M- and R-modes result in 15 unique space groups.
We display the group-subgroup relationships and \qty{0}{\kelvin} formation energies for \ce{BaZrS3} in \autoref{fig:Glazer_space}c.
As expected, \hmn{Pm-3m} is the highest energy phase relative to the \hmn{Pnma} ground state. 
\hmn{I4/mcm} is \SI{4.7}{\milli\electronvolt\per\atom} above the ground state, indicating that it may form as a higher temperature phase.
The comprehensive mapping across all possible structures ensures that we include all phases that might be formed at high temperatures in our training data for the machine-learning interatomic potential. 

Connecting lines in \autoref{fig:Glazer_space}c indicate group-subgroup relationships between structures.
In Landau theory, this relationship is necessary (but not sufficient) for structures connected through second-order (continuous) phase transitions.\cite{landau2013statistical}
Dashed lines indicate that, despite sharing a group-subgroup relationship, the phase transition must be first-order (discontinuous) in Landau theory \cite{stokes1984group}.

\begin{figure}
    \centering
    \includegraphics[width=\linewidth]{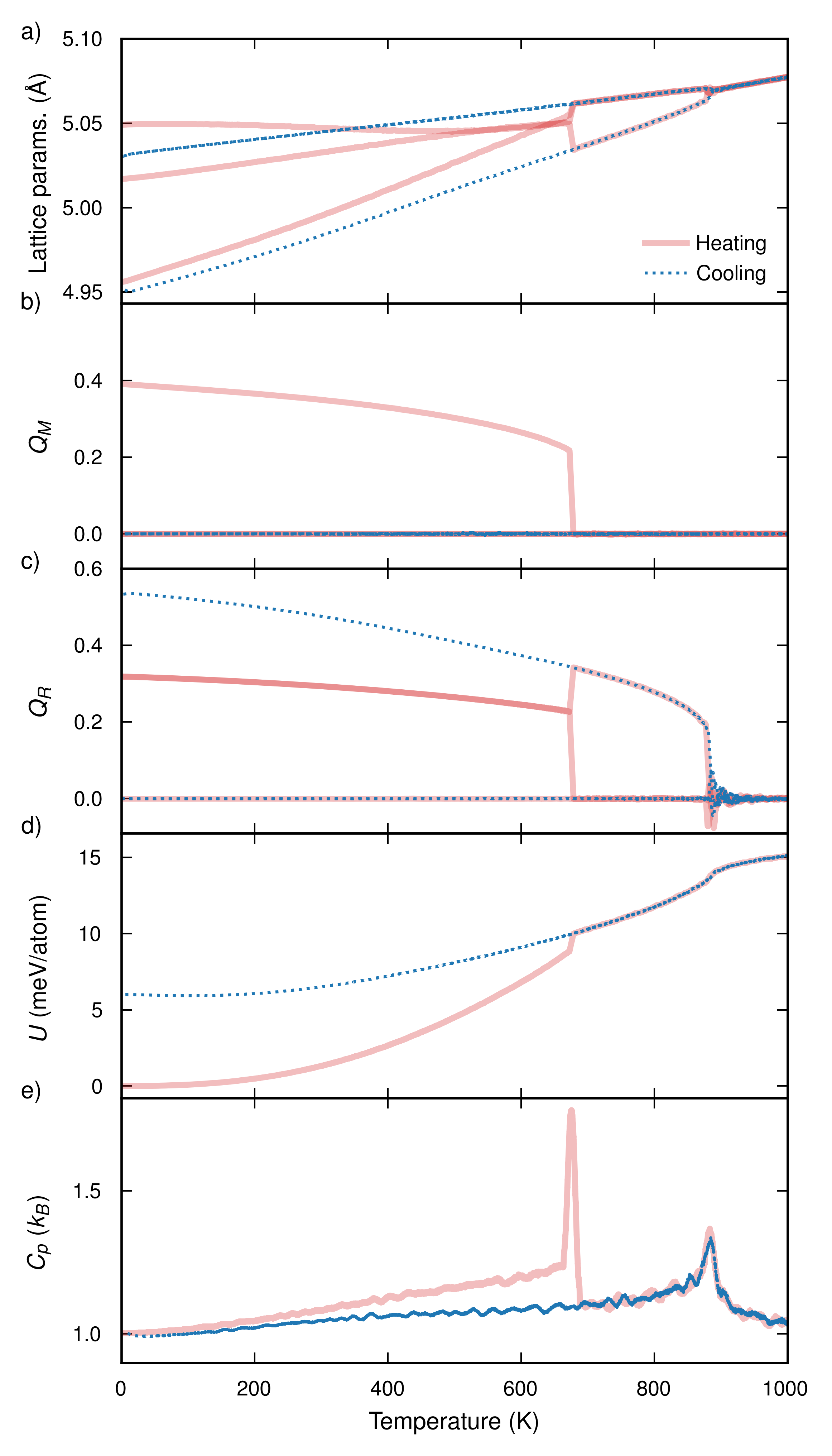}
    \caption{Properties of the \ce{BaZrS3} perovskite from cooling (blue) and heating (red) simulations: (a) pseudo-cubic lattice parameters; (b) M-mode and (c) R-mode amplitudes ($Q_\text{M}$, $Q_\text{R}$); (d) energies (U); (e) heat capacities ($C_p$).
    Energies are shown relative to the \hmn{Pnma} ground state energy at \qty{0}{\kelvin} with the equipartition energy ($\frac{3}{2}$ k$_B$T) subtracted. 
    The heat capacity is obtained by calculating the numerical derivative of the energy with respect to temperature, $C_p = dU/dT$ and is reported per degree of freedom in the system. 
    The simulation timescale is \qty{200}{\nano\second}. 
    All quantities are averaged over a time period of \qty{0.8}{\nano\second}.
    }
    \label{fig:phase_transitions}
\end{figure}

In \autoref{fig:phase_transitions}, we plot properties observed and derived from \gls{md} simulations spanning \qtyrange{0}{1200}{\kelvin} and with no applied pressure. 
When heating the low-temperature \hmn{Pnma} structure there are phase transitions at \qty{650}{\kelvin} and \qty{880}{\kelvin}. 
The transition at \qty{650}{\kelvin} is accompanied by a discontinuous and sharp change in lattice parameters (\autoref{fig:phase_transitions}a). 
Two of the lattice parameters become equal, indicating an orthorhombic to tetragonal transition. 
In contrast, the transition at \qty{880}{\kelvin} is gradual and continuous. 
All three lattice parameters become equal, indicating a tetragonal-to-cubic transition.

In \autoref{fig:phase_transitions}b,c we show projections of the M and R phonon modes on structures sampled from the simulation. 
From \qtyrange{0}{650}{\kelvin} one M-mode and two R-modes are active (have a non-zero amplitude).
This tilt pattern is described by $a^+ b^- b^-$ in Glazer notation and corresponds to a structure in the \hmn{Pnma} space group (see \autoref{fig:Glazer_space} and the associated discussion). 
From \qtyrange{650}{880}{\kelvin}, only one R-mode is activated corresponding to the tetragonal \hmn{I4/mcm} phase ($a^0 a^0 c^-$).
Above \SI{900}{\kelvin}, no modes are activated indicating the existence of a cubic \hmn{Pm-3m} phase ($a^0 a^0 a^0$). 

A sharp discontinuity is observed in energy (\autoref{fig:phase_transitions}d) and heat capacity (\autoref{fig:phase_transitions}e) at \qty{650}{\kelvin}. 
The \qty{1}{\milli\electronvolt\per\atom} energy change is the latent heat associated with a first-order phase transition and is comparable to that observed in other perovskites \cite{fransson2023phase, fransson2023revealing}.
At \qty{880}{\kelvin} a continuous change in energy is observed, typical of second-order phase transitions, resulting in a broader, less pronounced peak in the heat capacity.
We conclude that there is a first-order \hmn{Pnma}-to-\hmn{I4/mcm} transition at \qty{650}{\kelvin}, and a second-order \hmn{I4/mcm}-to-\hmn{Pm-3m} transition at \qty{880}{\kelvin}.
These observations are consistent with the group-subgroup analysis presented in \autoref{fig:Glazer_space}. 
The \hmn{Pnma} phase does not share a group-subgroup relationship with \hmn{I4/mcm}, necessitating a first-order phase transition.
In contrast, \hmn{P4/mbm} is a subgroup of the \hmn{Pm-3m} phase, so can be accessed through a second-order transition.

In the cooling runs, we start from the high-temperature \hmn{Pm-3m} structure and reproduce the heating behavior for the second-order transition at \qty{880}{\kelvin}. 
Significant hysteresis is observed for the first-order phase transition at \qty{650}{\kelvin} as the system cannot overcome the free energy barrier required to form the orthorhombic phase.
Due to the stochastic nature of \gls{md} simulations, we do recover the orthorhombic phase in some of the cooling runs (\autoref{fig:cooling-run-ortho}). 
Hysteresis in simulations describing a first-order transition has been observed and discussed in previous studies \cite{zhong1995first,fransson2023revealing, fransson2023phase}.

\label{sec:phase_diagram}
 \begin{figure}
     \centering
     \includegraphics[width=\linewidth]{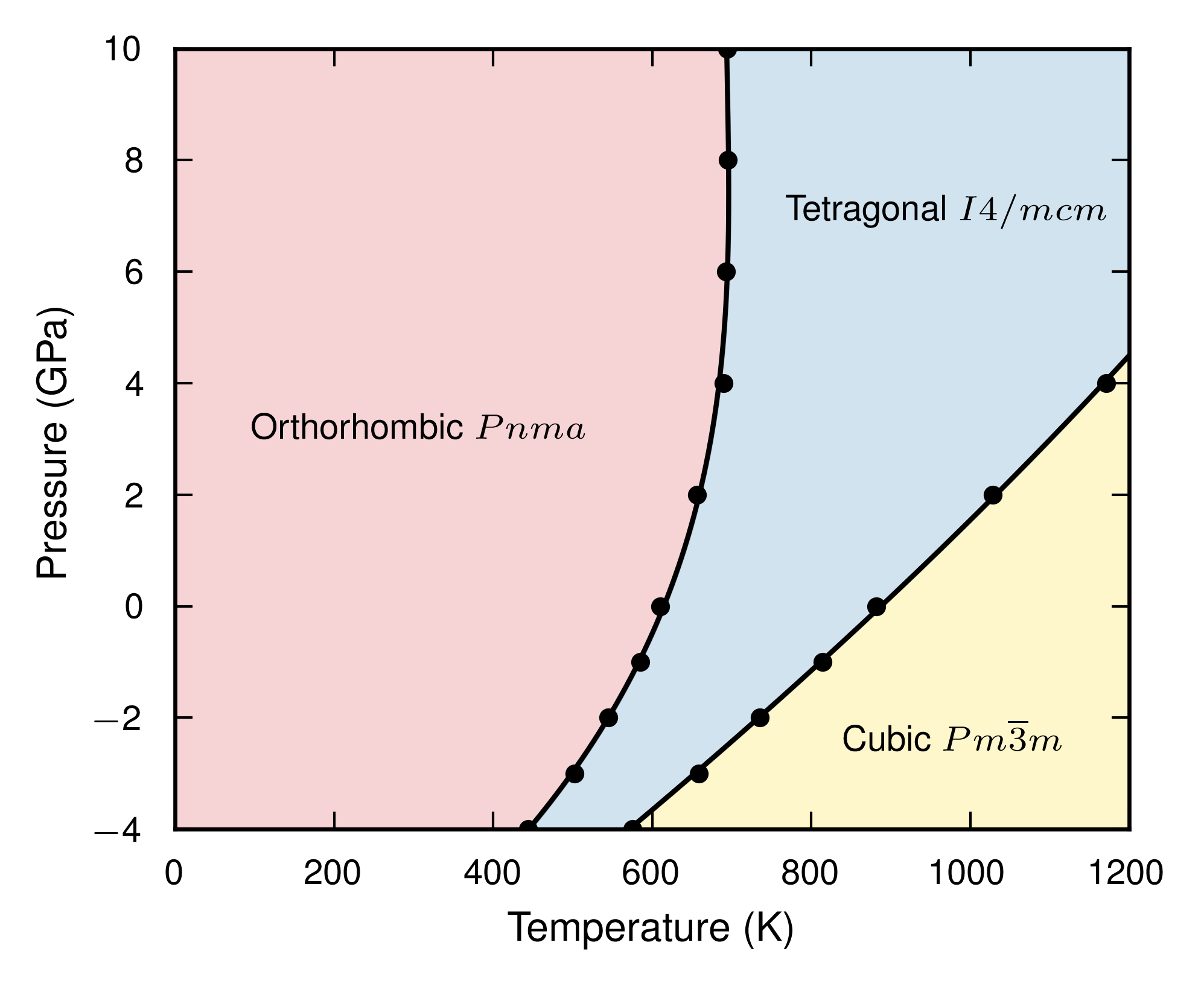}
     \caption{Phase diagram of \ce{BaZrS3} as a function of pressure and temperature.
     To predict the first-order \hmn{Pnma}-to-\hmn{I4/mcm} phase transition temperatures we use thermodynamic integration to calculate free energies.
     The second-order \hmn{I4/mcm}-to-\hmn{Pm-3m} phase transition temperatures are calculated from heating runs (\autoref{fig:phase_transitions}.)
     }
     \label{fig:phase_diagram}
 \end{figure}

In many materials transitions to new phases can be induced through applied pressure or strain.
In \autoref{fig:phase_diagram}, we present the \ce{BaZrS3} pressure-temperature phase diagram across \qtyrange{-4}{10}{\giga\pascal} and \qtyrange{0}{1000}{\kelvin}. 
Negative pressures correspond to triaxial tensile strain. Whilst this is difficult to realise experimentally, tensile strain in a plane can be produced through coherent interface formation with a suitably matched substrate \cite{Choi2024strain}.
As such, \autoref{fig:phase_diagram} indicates the range of polymorphs which might be accessed through interface engineering in a device stack.
To accurately predict the first-order \hmn{Pnma}-to-\hmn{I4/mcm} phase transition temperatures we use thermodynamic integration to calculate free energies.
This still describes the full anharmonicity of the material but avoids the kinetic limitations of a cooling/heating simulation which must overcome the first-order transition barrier \cite{fransson2023revealing}.  

The higher symmetry phases are stabilised with increasing temperature or decreasing pressure.
This indicates that the \ce{ZrS6} octahedra are relatively rigid, with volume expansion driven through decreased octahedral tilting;\cite{angel2005general} see section on compressibility in the SI for further discussion.
At zero pressure the \hmn{Pnma}-to-\hmn{I4/mcm} phase transition temperature is \qty{610}{\kelvin}. 
For comparison, the phase transition temperature in the harmonic approximation is \qty{243}{\kelvin} (\autoref{sfig:harmonic_approx}).
Above \qty{4}{\giga\pascal}, the transition temperature saturates at \qty{690}{\kelvin}. 
Below \qty{400}{\kelvin} there are no structural changes between \qtyrange{-4}{10}{\giga\pascal}.
This is in agreement with a previous \gls{dft} study at \qty{0}{\kelvin} and Raman measurements across the pressure range \qtyrange{0}{8.9}{\giga\pascal} \cite{Rong2022ab,gross2017stability}.

Our simulations show that \ce{BaZrS3} forms in the \hmn{Pnma} structure at room temperature.  
This observation is supported by experimental characterisation in ambient conditions and computational predictions at \qty{0}{\kelvin} \cite{niu2018thermal, gross2017stability, sopiha2022chalcogenide, Agarwal2025from}.
A recent computational study predicts that the polar \hmn{Pna2_1} phase is \qty{0.05}{\milli\electronvolt} per formula unit more stable than \hmn{Pnma} at \qty{0}{\kelvin} and \qty{0}{GPa} \cite{yaghoubi2024exotic}.
This instability has been observed across a variety of oxide perovskites in the orthorhombic phase \cite{Scott2024universal}.
For \ce{BaZrS3} the small \qty{0.05}{\milli\electronvolt} energy difference follows small differences in atomic coordinates, with an extremely tight symmetry tolerance of \qty{0.003}{\angstrom} required to differentiate the phases.

A multimodal study combining synchrotron \gls{xrd}, Raman spectroscopy, optical measurements and thermal analysis as a function of temperature identified three polymorphs when \ce{BaZrS3} is heated in air \cite{Jaiswal2024High}. 
Rietveld analysis of the synchrotron powder X-ray diffraction patterns showed the \hmn{I4/mcm} phase to be stable above \qty{770}{\kelvin} and the \hmn{Pnma} phase to be stable below \qty{570}{\kelvin}.
From \qtyrange{570}{770}{\kelvin} indirect observations suggest that \hmn{Cmcm} coexists as a minority phase.
Despite including the \hmn{Cmcm} phase in our training data, our simulations do not predict \hmn{Cmcm} as a stable intermediate phase.
In fact, at \qty{0}{\kelvin} our \gls{dft} calculations show that this phase is kinetically unstable and relaxes to the higher symmetry \hmn{I4/mcm} phase (\autoref{tab:mode_projection_DFT}). 

In a separate study from Bystrický \textit{et al.}, temperature-dependent \gls{xrd} data from a non-synchrotron source also indicated an orthorhombic-to-tetragonal phase transition at \qty{770}{\kelvin} \cite{bystricky2024thermal, jaykhedkar2023temperature}.
Full structure refinements were not presented and the measurements were partially obstructed through oxidation.
According to the analysis of that data above \qty{770}{\kelvin}, the two unique lattice parameters converge \cite{jaykhedkar2023temperature}, which we also observe whilst approaching the second-order \hmn{I4/mcm}-to-\hmn{Pm-3m} transition in our \gls{md} simulation (\autoref{fig:phase_transitions}a).

\begin{figure}
    \centering
    \includegraphics[width=\linewidth]{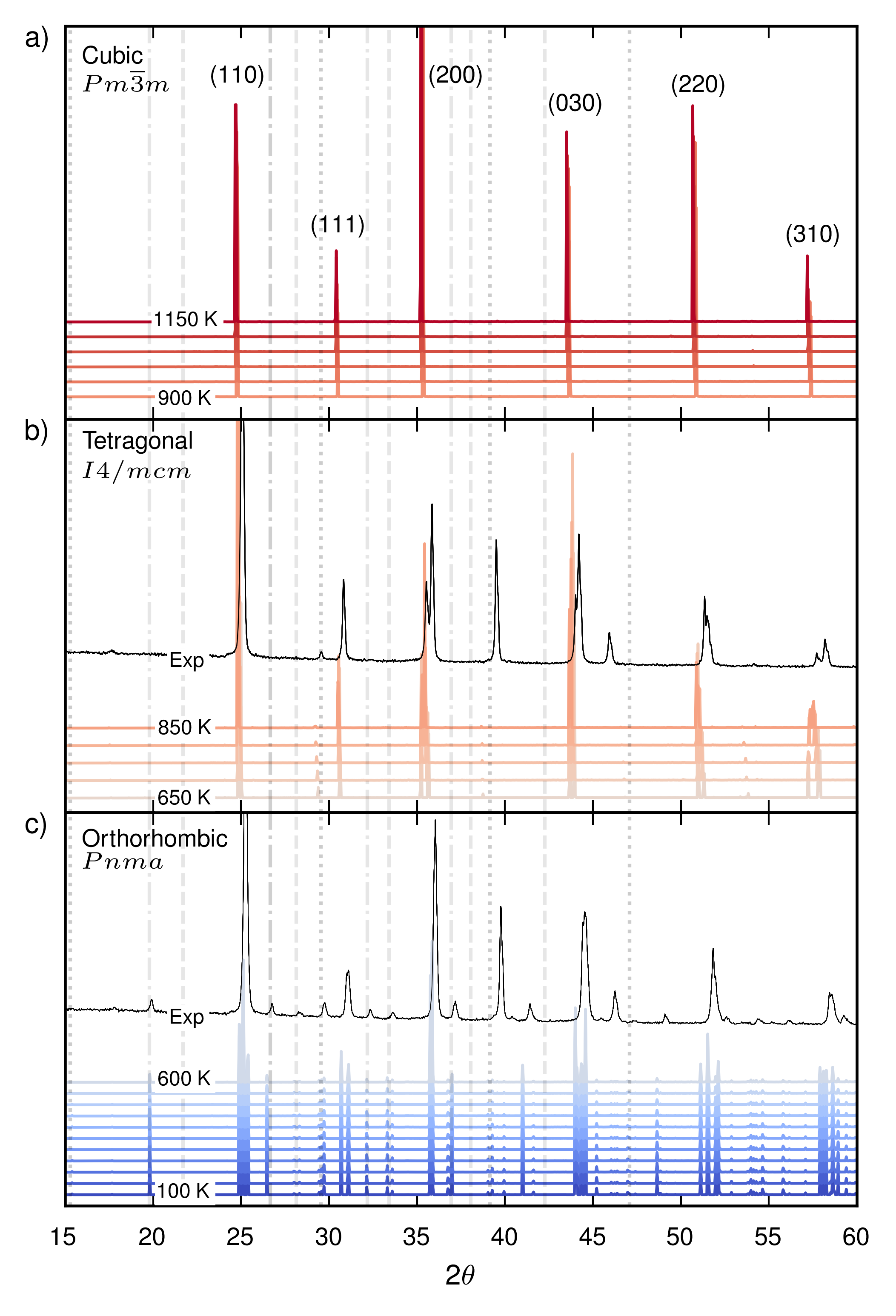}
    \caption{X-ray diffraction pattern evaluated for three 
    \ce{BaZrS3} polymorphs. The temperature ranges from \qtyrange{100}{1150}{\kelvin} in intervals of \qty{50}{\kelvin}. All simulations are at \qty{0}{\pascal}. A Cu K$\alpha$ value of \qty{1.5406}{\angstrom} was used for the $q$ to $\theta$ conversion.
    Cubic \hmn{Pm-3m} peaks are indexed. 
    Superlattice peaks at half integer planes up to the third Brillouin zone are indicated with vertical lines. The R-, M- and X-point distortions are represented with dotted, dashed and dash-dotted lines, respectively. For comparison, \autoref{fig:XRD_full} displays the static structure factor up to the fifth Brillouin zone.
    Experimental XRD data for the orthorhombic (\qty{303}{\kelvin}) and tetragonal (\qty{923}{\kelvin}) phases from Ref.~\citenum{bystricky2024thermal} are plotted in black. The peaks in the experimental data at \ang{39} and \ang{46} are associated with the Pt strip used for heating the sample.}
    \label{fig:XRD}
\end{figure}

We present the temperature dependent X-ray diffraction pattern in \autoref{fig:XRD}. 
At high temperature, the characteristic peaks of a cubic perovskite are clearly identified (\autoref{fig:XRD}a).
Below \qty{900}{\kelvin}, we observe peak splittings of the cubic diffraction lines. 
From \qtyrange{650}{900}{\kelvin} the largest splitting corresponds to a \textit{h00} reflection (200), and the \textit{hhh} reflection (111) remains a singlet, indicating a tetragonal distortion \cite{howard2005structures}.

Experimental \gls{xrd} data from Bystrický \textit{et al.} is displayed in \autoref{fig:XRD}b for comparison against our predictions \cite{bystricky2024thermal}.
The structure was refined in the tetragonal space group \hmn{I4_1/acd} which, to the best of our knowledge, has not been previously reported for an \ce{ABX3} perovskite.
We find good agreement between the experimental results and our prediction for the higher-symmetry \hmn{I4/mcm} phase.
R-mode activation produces a doubling of the unit cell along one axis and the appearance of a superlattice peak at a half-integer plane.
This can be seen in both the experimental and simulation data at $2\theta\approx\ang{29}$.

Below \qty{600}{\kelvin}, an abrupt change is observed due to the first-order phase transition into the orthorhombic phase (\autoref{fig:XRD}c).
M-point distortions result in the appearance of additional superlattice peaks at \ang{27} and \ang{33} alongside further peak splitting.
Superlattice peaks associated with X-mode distortions also appear at \ang{20}, \ang{26}, \ang{32} and \ang{37}. 
The same behavior is observed in the experimental \gls{xrd} data measured at \qty{303}{\kelvin}.\cite{bystricky2024thermal}
Decomposition of the X-ray diffraction intensity (\autoref{fig:atomic-weighted-XRD}) shows that some X-point peaks are associated with off-centering of the Ba species. 
A-site cation off-centering is frequently observed when R-point and M-point distortions operate in tandem.\cite{Glazer1975simple}

Transitions between structurally similar perovskite phases are not always discernible in Raman spectra, as the peak splitting can be less than the peak broadening resulting from thermal fluctuations or higher-order scattering \cite{Cohen2022, menahem2023disorder, rosander2024untangling}.
The spectral energy densities of the \hmn{I4/mcm} and \hmn{Pm-3m} phases demonstrate that there is considerable phonon broadening at elevated temperatures (\autoref{fig:SED_I4_mcm_Pm3m}). 
Whilst Ye \textit{et al.} reported that there is no indication of a first-order phase transition between \qtyrange{10}{875}{\kelvin}\cite{ye2024differing},
Jaiswal \textit{et al.} found the number of Raman peaks to decrease with increasing temperature, indicative of a phase transition to a higher-symmetry structure \cite{Jaiswal2024High}.
Our simulated Raman spectra in \autoref{sfig:Raman_Pnma} and \autoref{sfig:Raman_I4_mcm} demonstrates that there is significant peak overlap between the \hmn{Pnma} and \hmn{I4/mcm} phases.
Our spectra also reproduces the two most pronounced changes with temperature from Jaiswel \textit{et al.}: removal of the $A^6_g$ peak and a significant shift in the $B^6_g$ peak position.

Experimental characterisation of the high-temperature \hmn{Pm-3m} phase is hindered by oxidation which leads to the formation of \ce{BaSO4}, \ce{ZrO2} and \ce{SO2}. Differential scanning calorimetry and thermogravimetric analysis show that \ce{BaZrS3} is stable in air up to \qty{920}{\kelvin}, with complete conversion to the oxidised products at \qty{970}{\kelvin} \cite{Jaiswal2024High, niu2018thermal}.

In conclusion, chalcogenide perovskites, in particular \ce{BaZrS3}, show great potential for applications in optoelectronic and thermoelectric technologies. 
However several aspects of fundamental material behavior, including polymorphic phase transitions, have not yet been explored in detail.
In addition, experimental characterisations of the structure through Raman spectroscopy and X-ray diffraction give conflicting results.
We address this problem by developing a machine-learning interatomic potential for \ce{BaZrS3} trained on hybrid \gls{dft} calculations. 
This is used to run \gls{md} simulations across a wide range of temperatures and pressures.

The structural and thermodynamic properties derived from heating simulations reveal a series of transitions from orthorhombic \hmn{Pnma}-to-tetragonal \hmn{I4/mcm}-to-cubic \hmn{Pm-3m} with increasing temperature. 
Whilst this sequence of structures---from the low-symmetry \hmn{Pnma} phase to the high-symmetry \hmn{Pm-3m} phase---is commonly observed in perovskite materials, to the best of our knowledge this is the first report for \ce{BaZrS3}.
There is no evidence for additional transitions beyond these before melting.
The predicted character of each transition---first-order \hmn{Pnma}-to-\hmn{I4/mcm} and second-order \hmn{I4/mcm}-to-\hmn{Pm-3m}---is in agreement with those allowed by group-subgroup relationships. 

Both phase transitions occur above \qty{600}{\kelvin}, which agrees with experimental characterisation showing \ce{BaZrS3} is stable in the orthorhombic \hmn{Pnma} phase at ambient temperature and pressure. 
In addition, the calculated Raman spectra and temperature-dependent X-ray diffraction pattern align well with experimental data, supporting our prediction of an orthorhombic-to-tetragonal phase transition and validating our overall approach. 
The second-order transition at \qty{880}{\kelvin} is more difficult to characterise due to the concurrent high-temperature oxidation processes; further experimental studies in an inert atmosphere are required for confirmation.

It is possible that \ce{BaZrS3} samples grown at high temperature may include mixtures of polymorphs.
Future work might more fully consider polymorph mixing, alongside the impact of octahedral tilting on the thermal and optoelectronic properties of \ce{BaZrS3}. 
We note that the formation of ternary Ruddlesden-Popper phases \ce{Ba_{n+1}Zr_{n}S_{3n+1}} has been considered elsewhere in the literature \cite{kayastha2023high,kayastha2024model,Pradhan2024emergence}.
When formed these are likely to have a greater impact on material properties through disruption of the 3D octahedral framework.

\section*{Data Availability Statement}

The \gls{nep} models generated in this study are openly available via Zenodo at \url{https://doi.org/10.5281/zenodo.14229468}. The raw data from the \gls{dft} calculations have been uploaded to the NOMAD repository \url{https://dx.doi.org/10.17172/NOMAD/2024.11.25-2}.
A separate repository is also hosted at 
\url{https://github.com/NU-CEM/2024_BaZrS3_Phase_Transitions} with Python code available to reproduce the figures and analysis.

\section*{Supporting Information}

The Supporting Information file contains further details on Methods and NEP model validation, phonon mode projections for the Glazer tilt structures, results from a cooling run at \qty{-1}{\giga\pascal}, temperature-dependent phonon dispersions from molecular dynamics, harmonic free energies and Raman spectra, a discussion on perovskite bond compressibility, and temperature-dependent static structure factors.

\section*{Acknowledgements}

P.K. is grateful for funding through the Turing Scheme, which facilitated a research visit to Chalmers University of Technology. P.K. also acknowledges support from the UK Engineering and Physical Sciences Research Council (EPSRC) CDT in the Renewable Energy Northeast Universities (ReNU) for funding through EPSRC Grant EP/S023836/1. This work used the Oswald High-Performance Computing Facility operated by Northumbria University (UK). Via our membership in the UK’s HEC Materials Chemistry Consortium, which is funded by EPSRC (EP/X035859), this work used the ARCHER2 UK National Supercomputing Service. We are grateful to the UK Materials and Molecular Modelling Hub for computational resources, which is partially funded by EPSRC (EP/T022213/1, EP/W032260/1, and EP/P020194/1). 

This work has also been supported by the Swedish Research Council (Nos. 2020-04935 and 2021-05072), and the Chalmers Initiative for Advancement of Neutron and Synchrotron Techniques. 
Some of the computations were enabled by resources provided by the National Academic Infrastructure for Supercomputing in Sweden (NAISS) at C3SE, partially funded by the Swedish Research Council through grant agreement no. 2022-06725, as well as the Berzelius resource provided by the Knut and Alice Wallenberg Foundation at NSC.

We thank Fredrik Eriksson, Kostiantyn Sopiha and Florian Knoop for discussion related to this study. We thank Milan S\'ykora and team for sharing the experimental X-ray Diffraction data displayed in Figure 4.

\bibliography{references}

\end{document}


\maketitle

\tableofcontents{}

\newpage
\section{Methods}
We constructed \gls{nep} models by employing the iterative strategy outlined in Ref.~\citenum{FraWikErh2023}.
The \textsc{gpumd} package in version 3.9.4 \cite{FanZenZha21, Fan22, FanWanYin22} was used to build the \gls{nep} model and run the \gls{md} simulations.
The \ase{} \cite{Larsen2017} and \calorine{} \cite{calorine} packages were used to prepare the training structures, set up \gls{md} simulations and post-process the results.

The initial training set contains strained primitive structures and rattled supercells of the cubic (\hmn{Pm-3m}), tetragonal (\hmn{I4/mcm}) and orthorhombic (\hmn{Pnma}) perovskite phases, alongside structures corresponding to each of the 15 unique perovskite tilt patterns as specified by Glazer cite{glazer1972classification}.
Random displacements were generated using the \hiphive{} package \cite{EriFraErh19}.
The training set also contains \num{92} Ruddlesden-Popper structures, which will be the subject of a future publication.

The initial model was trained using \gls{dft} data generated with the PBEsol exchange-correlation functional \cite{perdew2008restoring}.
For each structure in the training set, we calculated the formation energy (relative to the elemental phases), stress tensor and atomic force.
This model was then used to run \gls{md} simulations in the NPT ensemble, over a temperature and pressure range of \qtyrange{0}{1200}{\kelvin} and \qtyrange{-5}{20}{\giga\pascal} respectively, with varying supercell sizes, containing between 20 and 500 atoms.
Snapshots from the \gls{md} simulations were randomly selected and added to the training set, after which the model was retrained.

Finally, a higher accuracy training set was generated using \gls{dft} with the hybrid functional HSE06 \cite{krukau2006influence}.
Here we carried out single-point calculations on all \num{1187} training structures generated during the construction of the PBEsol-based model.
Stress tensors for structures containing more than 60 atoms were not evaluated as the memory requirements for these calculations were prohibitively large.
A comparison of models at the PBEsol level of theory shows that using a sub-set of stress tensors does not impact the predicted phase transition temperature or other properties of interest.


\subsection{Density Functional Theory calculations}
The training data for each \gls{nep} model was generated using \gls{dft} calculations to evaluate the formation energies (relative to elemental phases), stress tensors and forces.
These calculations were performed using the all-electron numeric atom-centered orbital code FHI-aims \cite{blum2009ab}.
FHI-vibes \cite{knoop2020fhi} was used for pre and post-processing of \gls{dft} data.  All \gls{dft} calculations used the $\textit{light}$ basis set and a Monkhorst-Pack $k$-point mesh with a minimum $k$-spacing of \qty{0.2}{\per\angstrom}.
For single-point calculations, the charge density was converged to an accuracy of \num{e-6}, forces to \qty{e-5}{\electronvolt\per\angstrom} and stresses to \qty{e-4}{\electronvolt\per\angstrom\cubed}. 
Geometry relaxations were carried out using the symmetry-constrained relaxation scheme as implemented in \ase{} \cite{Larsen2017}, until the maximal force component was below \qty{e-3}{\electronvolt\per\angstrom}.


Harmonic phonon dispersions at \qty{0}{\kelvin} and Helmholtz free energies were evaluated using the \textsc{phonopy} package \cite{togo2023first} with a \numproduct{2x2x2} supercell and a \qty{0.01}{\angstrom} displacement distance. 

\subsection{Molecular Dynamics}
\Gls{md} simulations were carried out using the \textsc{gpumd} software \cite{FanWeiVie2017} with a timestep of \qty{1}{\femto\second}.
The \ase{} \cite{Larsen2017} and calorine \cite{calorine} packages were used to set up the \gls{md} simulations and post-process the results.
Heating and cooling simulations were run in the NPT ensemble between \qtyrange{0}{1200}{\kelvin} for \qty{200}{\nano\second} using a supercell consisting of about \num{40960} atoms.
The potential energy and lattice parameters were recorded every \qty{100}{\femto\second} to discern phase transitions.

Free energy calculations were carried out via \gls{ti} using an Einstein crystal as reference Hamiltonian (also referred to as the Frenkel-Ladd method \cite{FreLad84}) as outlined in Ref.~\citenum{FreAstde16}.
The free energy of the system described by the \gls{nep}, $F_\text{NEP}$, is obtained from
\begin{align}
    F_\text{NEP} - F_\text{Ein} = \int_0^1 \left < \frac{\mathrm{d}H(\lambda)}{\mathrm{d}\lambda} \right >_{H} \mathrm{d}\lambda,
    \label{eq:lambda_integration}
\end{align}
where the integration is carried out over the Kirkwood coupling parameter $\lambda$ \cite{Kir35}, $F_\text{Ein}$ is the analytically known classical free energy of an Einstein crystal, and the Hamiltonian is $H(\lambda) = (1 - \lambda)$ $H_{\rm{Ein}}$ + $\lambda H_{\rm{NEP}}$, as implemented in \textsc{gpumd}.
Here, the ensemble average $\left < \ldots \right >_H$ is sampled using the Hamiltonian $H(\lambda)$.
The Gibbs free energy can then be obtained by $G = F + PV$.
\Gls{ti} simulations were run for \qty{0.05}{\nano\second} using a spring constant of \qty{4}{\electronvolt\per\angstrom\squared} using a supercell consisting of \num{23040} atoms.
These simulations were run in the NVT ensemble with lattice parameters obtained from NPT simulations.

The static structure factor, $S(\boldsymbol{q})$, is calculated from NVT simulations using the \textsc{dynasor} package \cite{FraSlaErhWah2021} as 
\begin{equation}
    S(\boldsymbol{q}) = \frac{1}{N} \left < \sum_{i } ^{N} \sum_{j} ^{N} \exp{\left [ i\boldsymbol{q}\cdot (\boldsymbol{r}_i(t) - \boldsymbol{r}_j(t)) \right ]} \right >,
\end{equation}
where $\boldsymbol{r}_i(t)$ is the position of atom $i$ at time $t$ and the sums run over all atoms.
To obtain the intensity measured in an X-ray experiment, $I(q)$, one must include the X-ray form factors as
\begin{equation}
    I(\boldsymbol{q}) = \frac{1}{N} \left < \sum_{i } ^{N} \sum_{j} ^{N} f_i(q) f_j(q) \exp{\left [ i\boldsymbol{q}\cdot (\boldsymbol{r}_i(t) - \boldsymbol{r}_j(t)) \right ]} \right >,
\end{equation}
where $f_i(q)$ is the q-dependent X-ray form factors, here taken from Ref.~\citenum{Waasmaier1995}.
The partial structure factors and intensities can be obtained by considering only specific atom types in the sums, see Ref.~\citenum{FraSlaErhWah2021}.
Lastly, we apply Bragg's law to convert the structure factor and intensity from q-space to $\theta$-space:
\begin{equation}
    \frac{sin(\theta)}{\lambda} = n\frac{q}{4\pi}
\end{equation}
where $\lambda$ is the wavelength of the incident X-ray beam used in experiment. 

We employ phonon mode projection to analyze and classify both relaxed structures and snapshots from \gls{md} simulations as done in Ref.~\citenum{FraRosEriRahTadErh23}.
The atomic displacements $\vec{u}$  can be projected on a mode $\lambda$, with the supercell eigenvector $\vec{e}_\lambda$, via
\begin{align*}
    Q_\lambda = \vec{u} \cdot \vec{e}_\lambda
\end{align*}
Here, the phonon supercell eigenvector of the $R$ and $M$ tilt modes are obtained with \textsc{phonopy} \cite{TogTan15}, and symmetrized such that each of the three degenerate modes corresponds to tilting around the x, y, and z direction respectively.
\newpage
\section{NEP model validation}
Our final training set consists of 1187 structures. Energies, forces, and stress tensors are evaluated using the HSE06 functional. 
Energies and forces are evaluated for all structures. However, due to large memory requirements, to evaluate stress tensors are only evaluated for structures with less than 60 atoms in the unit cell. 
Energy and force errors are reported for the entire training set of 1187 structures. Virial and stress errors are only reported for structures where the DFT stress tensors were evaluated. 

\begin{figure}[H]
    \centering
    \includegraphics[width=0.75\linewidth]{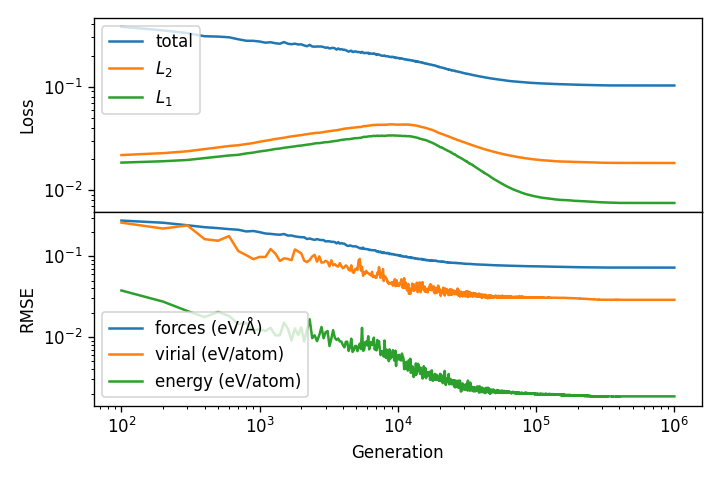}
    \caption{Loss curves for the HSE06 model}
    \label{sfig:loss_curves}
\end{figure}
\begin{figure}[H]
    \centering
    \includegraphics[width=\linewidth]{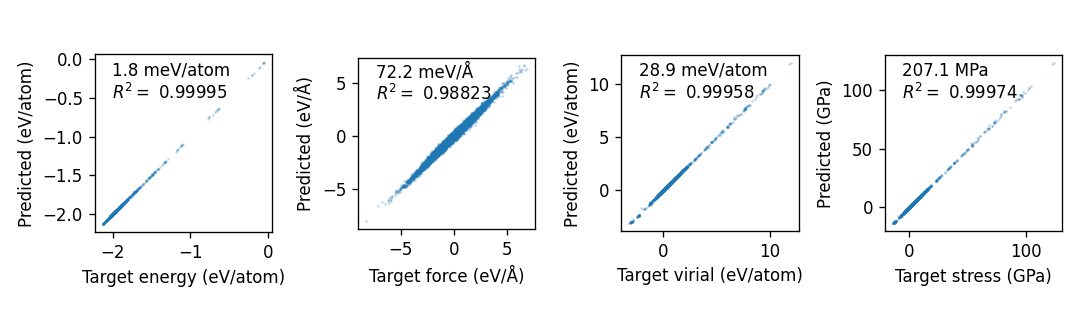}
    \caption{Parity plot for the HSE06 model. Structures for which the stress tensors have not been evaluated with \gls{dft} are not included in the virial and stress tensor parity plots.}
    \label{sfig:parity_plots}
\end{figure}
\newpage

\subsection{Comparison of phonon dispersion predicted from DFT and NEP}
The \qty{0}{\kelvin} harmonic phonon spectra for \ce{BaZrS3} in its three observed phases are displayed. The dashed black line corresponds to that generated with our NEP model, the solid blue is calculated using DFT-calculated forces. The largest discrepancies correspond to high-frequency modes strongly associated with the sulfur species.\cite{wu2023ultralow}

\begin{figure}[H]
    \centering
    \includegraphics[width=0.5\linewidth]{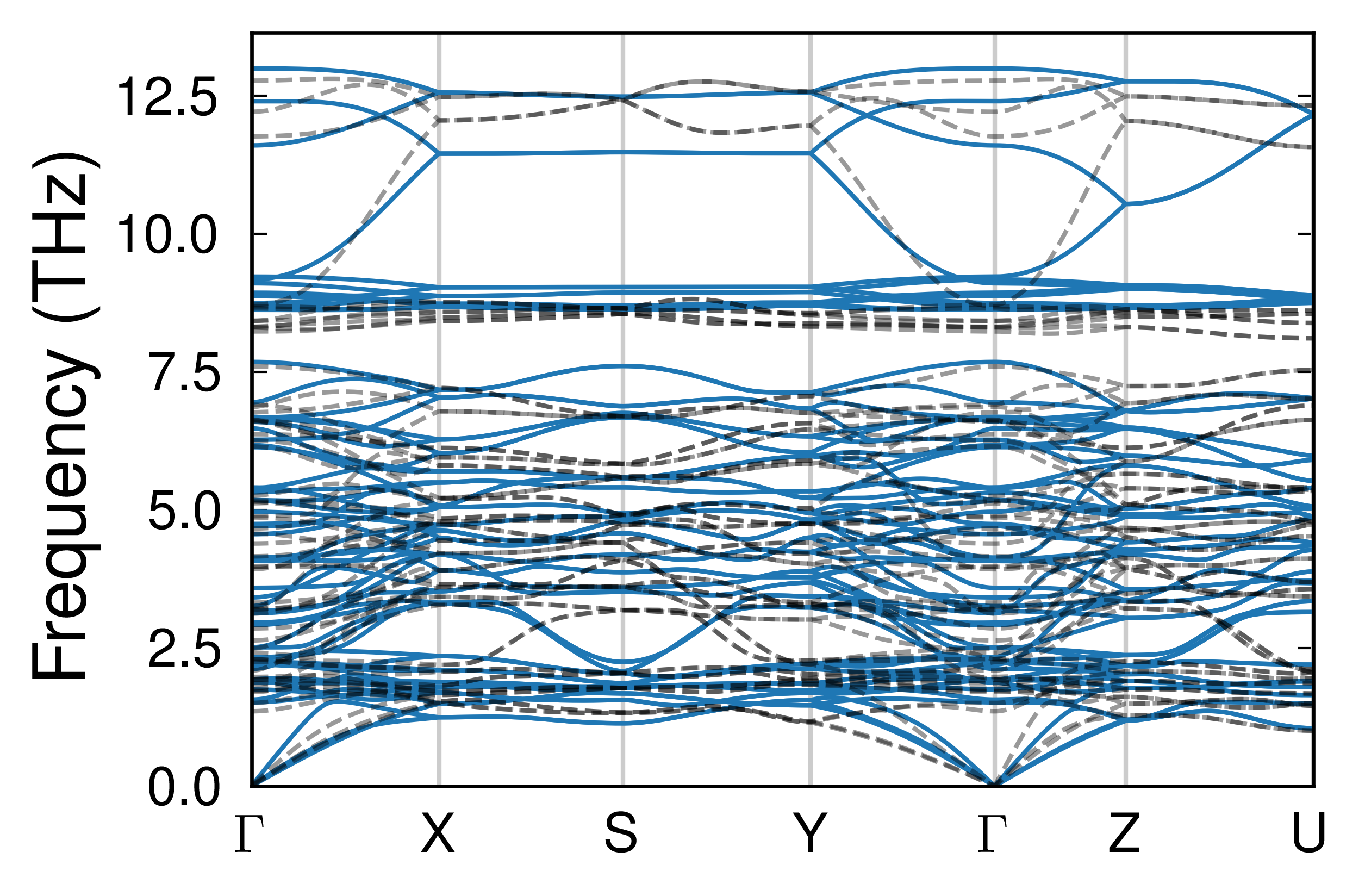}
    \caption{{Pnma} DFT (solid blue) vs NEP (dashed black) phonons }
    \label{sfig:Pnma_DFT_vs_NEP_phonons}
\end{figure}
\begin{figure}[H]
    \centering
    \includegraphics[width=0.5\linewidth]{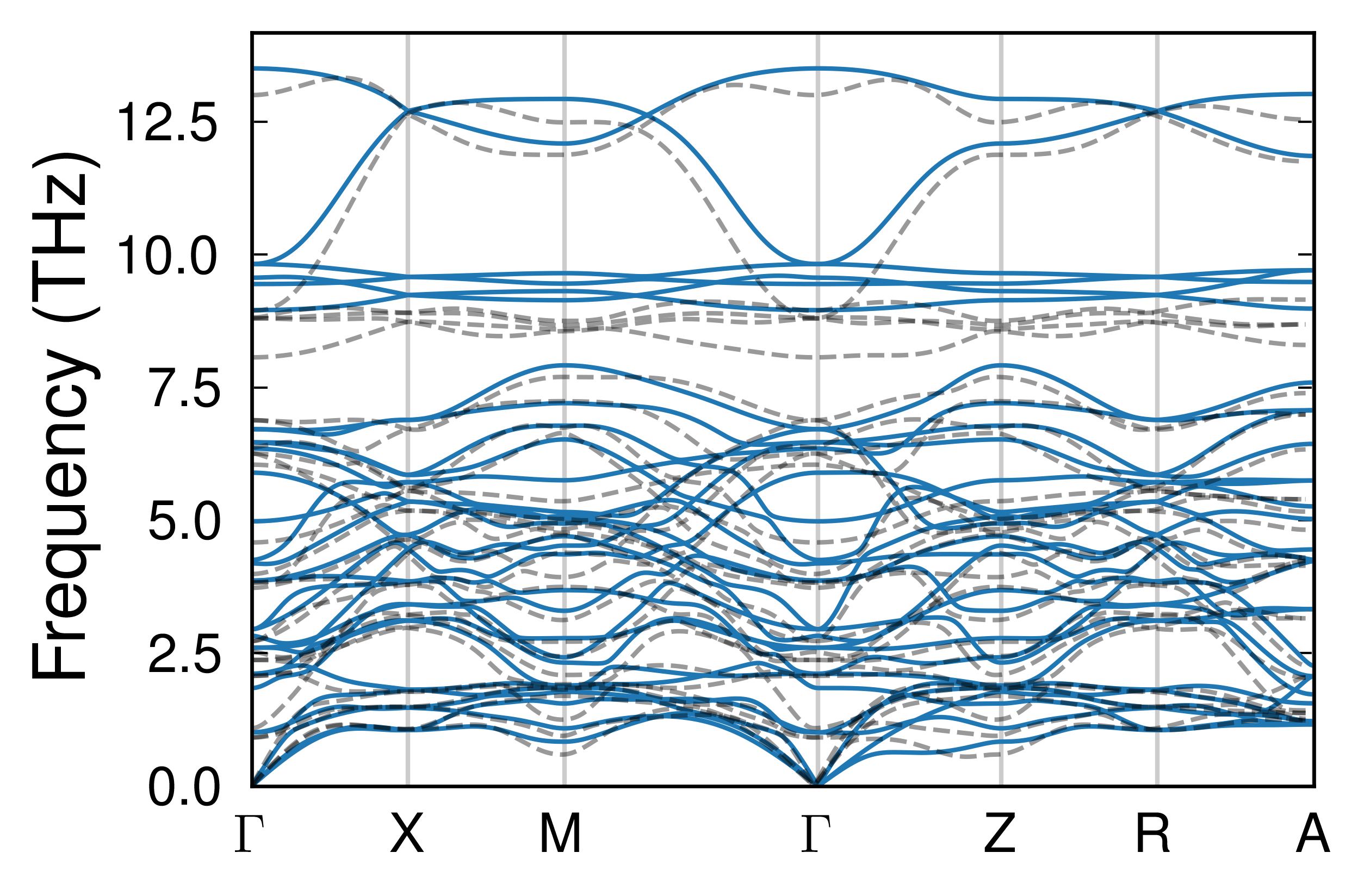}
    \caption{{I4/mcm} DFT (solid blue) vs NEP (dashed black) phonons}
    \label{sfig:I4_mcm_DFT_vs_NEP_phonons}
\end{figure}
\begin{figure}[H]
    \centering
    \includegraphics[width=0.5\linewidth]{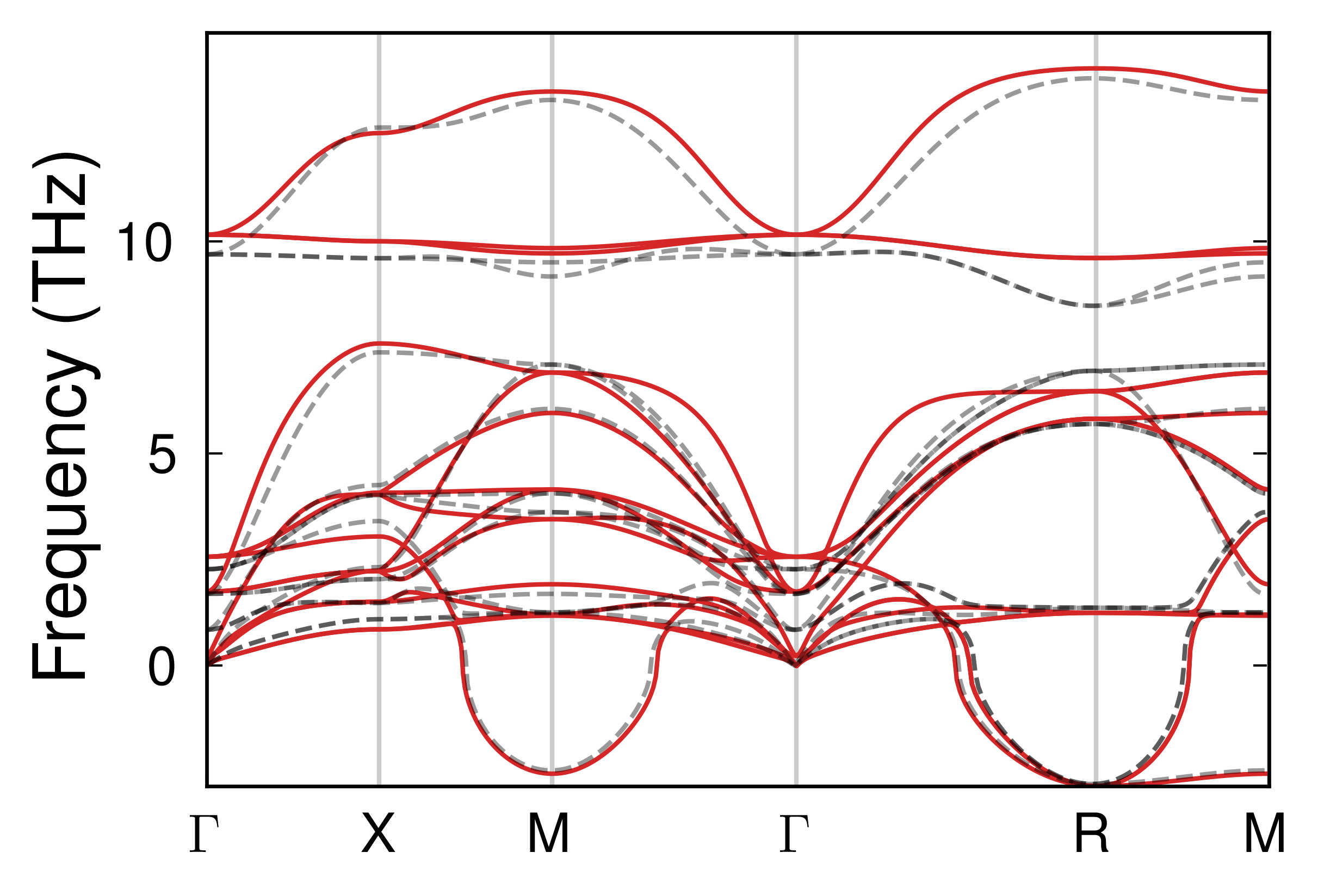}
    \caption{Pm$\Bar{3}$m DFT (solid red) vs NEP (dashed black) phonons}
    \label{sfig:Pm3m_DFT_vs_NEP_phonons}
\end{figure}

\newpage
\section{Mode projections for Glazer tilt structures}
In our comprehensive evaluation of all 15 Glazer-tilted structures we observe that \hmn{I4/mcm} and its subgroups (\hmn{P4_2/nmc}, \hmn{Cmcm}, \hmn{C2/m}, \hmn{C2/c}, \hmn{P-1}) have similar energies according to \gls{dft} calculations and the \gls{nep} model.
Our workflow for constructing these structures was to constrain the spacegroup symmetry using the \ase{}\cite{Larsen2017} \textsc{FixSymmetry} function.
Subgroup structures which are unstable at \qty{0}{\kelvin} relax to their supergroup phase with a small (but non-zero) phonon mode amplitude. The corresponding distortion lies within the tolerance factor of the symmetry-constrained relaxation (\qty{e-5}{\angstrom}).  
This small difference in structure results in energies that are similar but not exactly the same.
For example, the \hmn{Cmcm} $a^0b^+c^-$ structure relaxes to an \hmn{I4/mcm} $a^0b^0c^-$ equivalent structure with a small mode projection value for the $b^+$ tilt. This results in a small energy difference (\qty{0.2}{\milli\electronvolt}) between the two phases.

Similar observations can be made for the \hmn{P2_1/m} and \hmn{Pnma} phases, and the \hmn{Immm} and \hmn{P4/mbm} phases.

\begin{table}[H]
    \centering
    \begin{tabular}{lcccccc}
    Space group & M$_x$ & M$_y$ &  M$_z$ &  R$_x$ & R$_y$ & R$_z$ \\
    \hline
\hmn{Pm-3m} (221) & 0.000000 &  0.000000 & -0.000000 & -0.000000 &  0.000000 &  0.000000 \\
\hmn{I4/mcm} (140) & 0.000000 &  0.000000 &  0.000000 &  0.000000 &  0.000000 &  0.575006 \\
\hmn{P4/mbm} (127) & 0.000000 &  0.000000 &  0.548092 &  0.000000 &  0.000000 &  0.000000 \\
\hmn{Imma} (74) & 0.000000 &  0.000000 &  0.000000 &  0.000000 &  0.373962 &  0.373962 \\
\hmn{C2/m} (12) & 0.000000 &  0.000000 & 0.000000 & 0.000000 &  0.000030 &  0.575030 \\
\hmn{Cmcm} (63) & 0.000000 &  0.000002 &  0.000000 &  0.000000 & 0.000000 &  0.575030 \\
\hmn{I4/mmm} (139) & 0.000000 &  0.328544 &  0.328544 &  0.000000 &  0.000000 &  0.000000 \\
\hmn{R-3c} (167) & 0.000000 & 0.000000 &  0.000000 &  0.292519 &  0.292519 &  0.292519 \\
\hmn{C2/c} (15) & 0.000000 & 0.000000 & 0.000000 &  0.575000 & 0.000016 & 0.000016 \\
\hmn{P-1} (2) & 0.000000 & 0.000000 & 0.000000 &  0.574996 &  0.000037 &  0.000017 \\
\hmn{I4/mmm} (139) & 0.000000 &  0.328544 &  0.328544 &  0.000000 &  0.000000 &  0.000000 \\
\hmn{Pnma} (62) & 0.401595 & 0.000000 & 0.000000 & 0.000000 &  0.321830 &  0.321830 \\
\hmn{P2_1/m} (11) & 0.401610 &  0.000000 &  0.000000 &  0.000000 &  0.321820 &  0.321830 \\
\hmn{P4_2/nmc} (137) & 0.000001 &  0.000001 &  0.000000 & 0.000000 &  0.000000 &  0.575031 \\
\hmn{Im-3} (204) & 0.251797 &  0.251797 &  0.251797 &  0.000000 &  0.000000 &  0.000000 \\
\hmn{Immm} (71) & 0.548063 & 0.000001 & 0.000003 &  0.000000 &  0.000000 &  0.000000 \\
    \end{tabular}
    \caption{Mode projection on DFT relaxed geometries}
    \label{tab:mode_projection_DFT}
\end{table}

\begin{table}[H]
    \centering
    \begin{tabular}{lcccccc}
    Space group & M$_x$ & M$_y$ &  M$_z$ &  R$_x$ & R$_y$ & R$_z$ \\
    \hline
\hmn{Pm-3m} (221) & 0.000000 &  0.000000 &  0.000000 &  0.000000 &  0.000000 &  0.000000 \\
\hmn{I4/mcm} (140) & 0.000000 &  0.000000 &  0.000000 &  0.000000 &  0.000000 &  0.579234 \\
\hmn{P4/mbm} (127) & 0.000000 &  0.000000 &  0.542603 &  0.000000 &  0.000000 &  0.000000 \\
\hmn{Imma} (74) & 0.000000 &  0.000000 &  0.000000 & 0.000000 &  0.369080 &  0.369080 \\
\hmn{C2/m} (12) & 0.000000 & 0.000000 &  0.000000 &  0.000000 & -0.000028 &  0.579234 \\
\hmn{Cmcm} (63) & 0.000000 & 0.000000 &  0.000000 &  0.000000 &  0.000000 &  0.579234 \\
\hmn{I4/mmm} (139) & 0.000000 &  0.329391 &  0.329391 &  0.000000 &  0.000000 &  0.000000 \\
\hmn{R-3c} (167) & 0.000000 & 0.000000 &  0.000000 &  0.286208 &  0.286208 &  0.286208 \\
\hmn{C2/c} (15) & 0.000000 & 0.000000 & 0.000000 &  0.579234 & 0.000025 & 0.000025 \\
\hmn{P-1} (2) & 0.000000 & 0.000000 & 0.000000 &  0.579233 & 0.000045 & 0.000011 \\
\hmn{I4/mmm} (139) & 0.000000 &  0.329391 &  0.329391 &  0.000000 &  0.000000 &  0.000000 \\
\hmn{Pnma} (62) & 0.400823 &  0.000000 & 0.000000 &  0.000000 &  0.326056 &  0.326056 \\
\hmn{P2_1/m} (11) & 0.400821 &  0.000000 & 0.000000 &  0.000000 &  0.326061 &  0.326054 \\
\hmn{P4_2/nmc} (137) & 0.000000 & 0.000000 &  0.000000 &  0.000000 &  0.000000 &  0.579234 \\
\hmn{Im-3} (204) & 0.251378 &  0.251378 &  0.251378 &  0.000000 &  0.000000 &  0.000000 \\
\hmn{Immm} (71) & 0.542603 &  0.000000 & 0.000000 &  0.000000 &  0.000000 & 0.000000 \\
    \end{tabular}
    \caption{Mode projection on NEP relaxed geometries}
    \label{tab:mode_projection_NEP}
\end{table}
\newpage

\section{Recovery of the orthorhombic phase during cooling at \qty{-1}{\giga\pascal}}
Our cooling simulations in the main text do not recover the orthorhombic \ce{Pnma} phase formed at \qty{0}{\kelvin}.
This stems from the numerical limitations of our simulations; the short timescales considered make a first-order transition (with associated kinetic barrier) from the tetragonal phase implausible.
Here, we plot the heating and cooling runs at \qty{1}{\giga\pascal}, where the orthorhombic phase is recovered around \qty{560}{\kelvin}. 
This variation in our results is due to the stochastic nature of molecular dynamics simulations.
\begin{figure}[H]
    \centering
    \includegraphics[width=0.6
\linewidth]{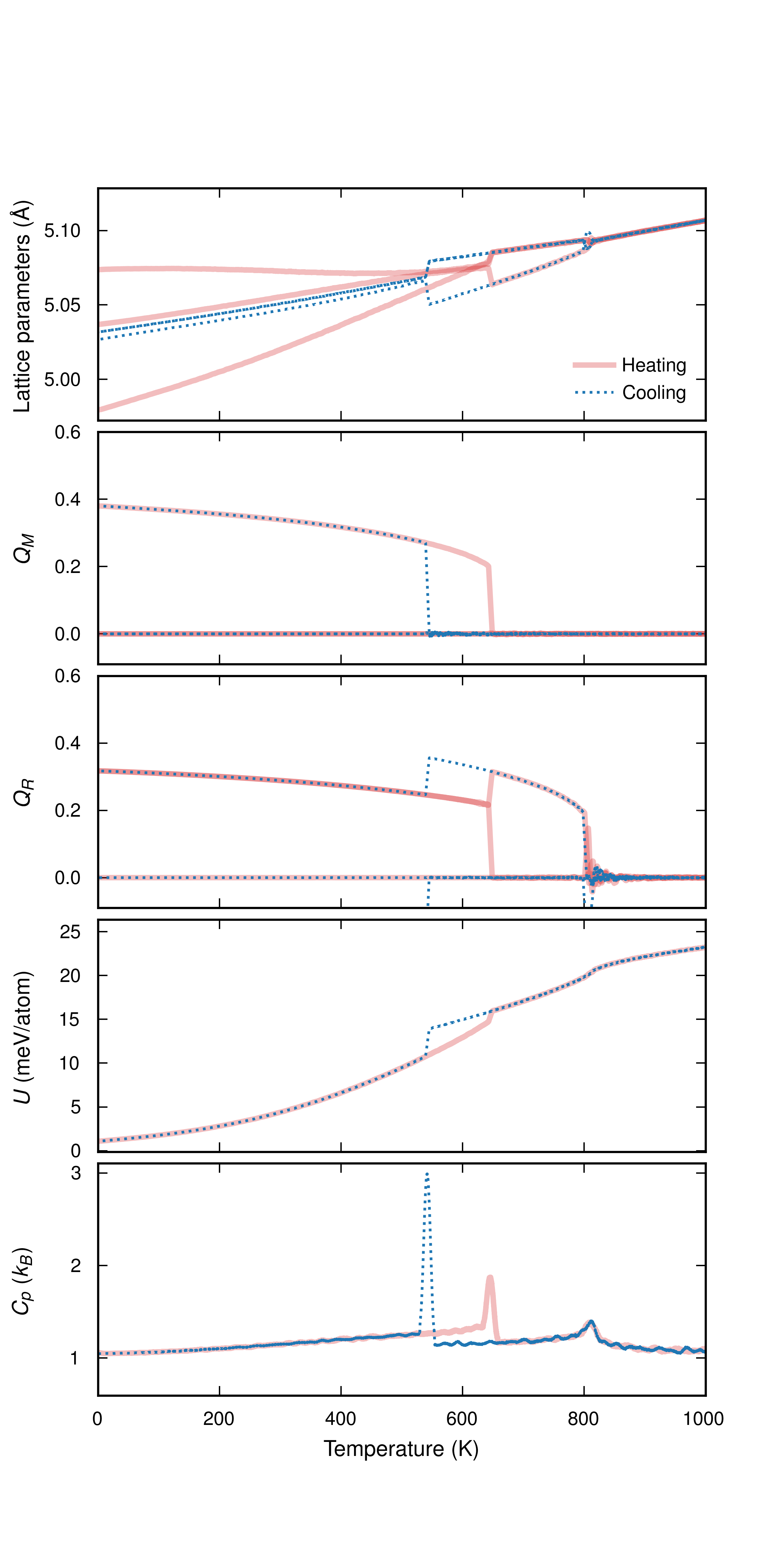}
    \caption{Heating and cooling runs at pressure \qty{-1}{\giga\pascal}. The orthorhombic \ce{Pnma} phase is recovered in the cooling run, unlike our simulations at \qty{0}{\pascal} (see the main text).}
    \label{fig:cooling-run-ortho}
\end{figure}
\newpage
\section{Finite-temperature phonons from molecular dynamics}
The \textsc{dynasor} package\cite{FraSlaErhWah2021} is used to calculate the spectral energy density of the \hmn{I4/mcm} and \hmn{Pm3m} phases at various temperatures.
\begin{figure}[H]
    \centering
    \includegraphics[width=\linewidth]{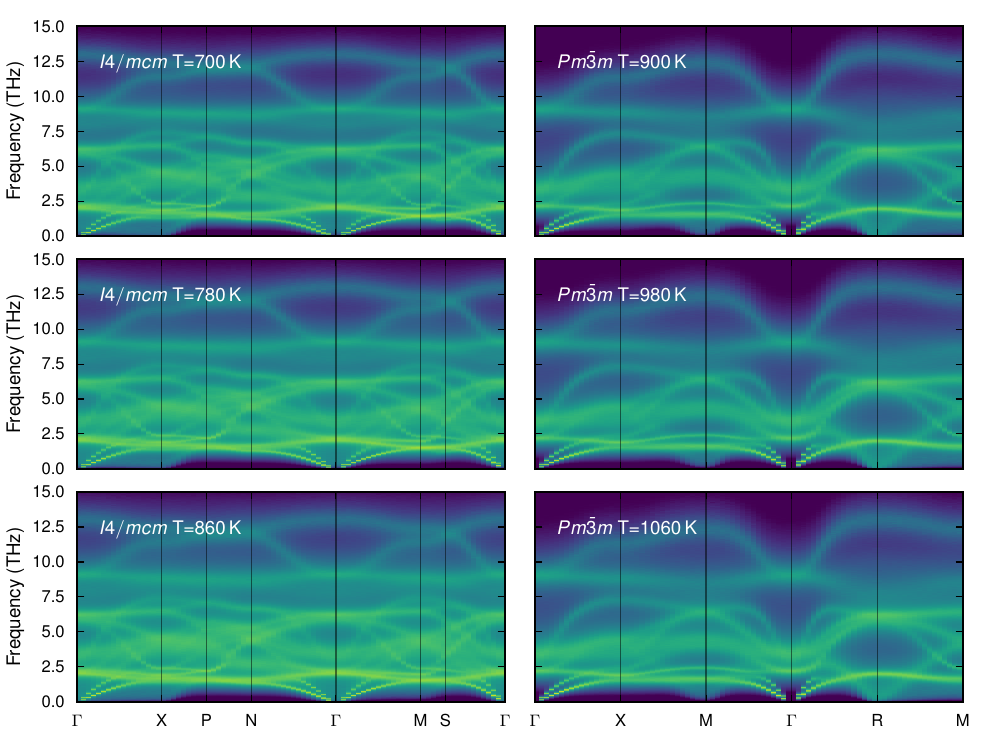}
    \caption{Spectral energy densities of the tetragonal \hmn{I4/mcm} and cubic \hmn{Pm-3m} phases of \ce{BaZrS3}.}
    \label{fig:SED_I4_mcm_Pm3m}
\end{figure}

\newpage
\section{Phase transition temperature using the harmonic approximation}
Helmholtz free energies of the orthorhombic (\hmn{Pnma}) and tetragonal (\hmn{I4/mcm}) phases are evaluated within the harmonic approximation. The phase transition temperature with DFT-calculated phonon frequencies is \qty{460}{\kelvin}, and with NEP-calculated frequencies it is \qty{243}{\kelvin}. The phase transition temperature predicted with our fully anharmonic molecular dynamics model (see main text) is  \qty{610}{\kelvin}.
\begin{figure} [H]
    \centering
    \includegraphics[width=0.5\linewidth]{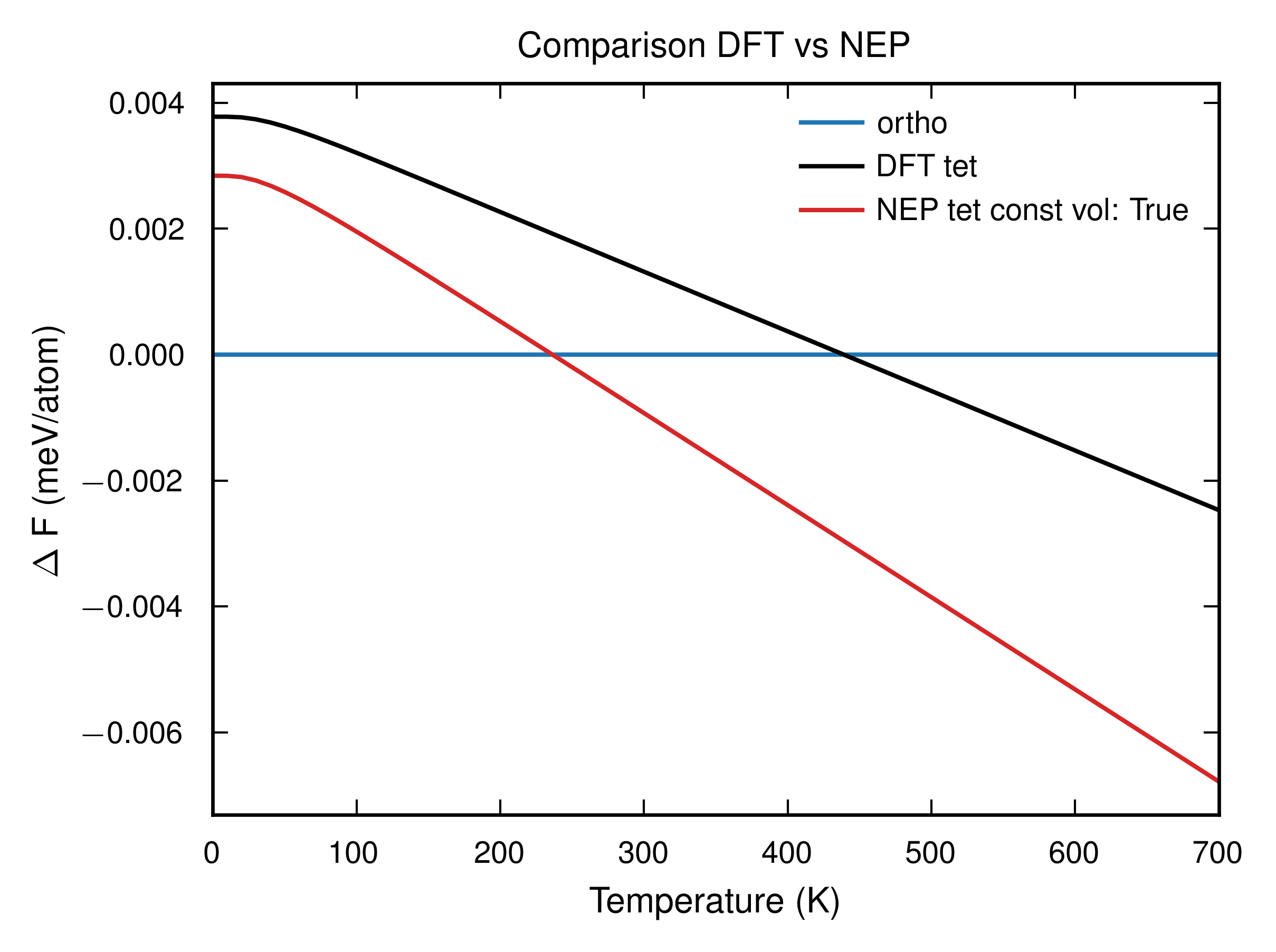}
    \caption{Helmholtz free energy difference between \hmn{Pnma} and \hmn{I4/mcm} phases ($\Delta F$), calculated under the harmonic approximation. }
    \label{sfig:harmonic_approx}
\end{figure}
\newpage
\section{Perovskite bond compressibility}
\label{snote:compressibility}
Perovskites with lower symmetry (more tilted) structures generally exhibit a series of octahedral tilt-driven phase transitions to higher symmetry (less tilted) structures with increasing temperature. 
The phase transition temperature $T_c$ can increase or decrease with pressure, $\frac{dT_c}{dP}>0$ or $\frac{dT_c}{dP}<0$ respectively.\cite{angel2005general}
For \ce{BaZrS3} we find $\frac{dT_c}{dP}>0$ (Fig 3 in the main text).
This behaviour can be rationalised by considering the relative compressibility of the A-X and B-X bonds.\cite{Zhao2004,angel2005general} 
If the B-X bonds forming the octahedra are relatively rigid then pressure must induce tilting to accommodate the volume reduction, resulting in the compression of the A-X bonds and a reduction in symmetry.
If the A-X bonds are relatively rigid then pressure must induce a reduction in the octahedral volume through compression of the B-X bond.
The ratio of the B-X and A-X bond compressibilities ($\frac{\beta_B}{\beta_A}$) can be used to indicate how the phase transition temperature will vary as a function of pressure: $\frac{dT_c}{dP}>0$ when $\frac{\beta_B}{\beta_A} < 1$, and $\frac{dT_c}{dP}<0$ when $\frac{\beta_B}{\beta_A} > 1$.
The parameter $M_i = \beta_i^{-1}$ is given by:
\begin{equation}
    M_i = \frac{R_i N_i}{B}\exp{\left (\frac{R_i - R_0}{B} \right )}
\end{equation}
where $N_i$ is the coordination number, $R_i$ average bond distance (to X-atoms), $B$ is a constant with value 0.37, and $R_0$ is the bond-valence parameter.\cite{Brown1985}
For \ce{BaZrS3} we obtain $\frac{\beta_B}{\beta_A} = \frac{M_A}{M_B} \approx 0.45$. This indicates that $\frac{dT_c}{dP}>0$, in agreement with our predicted phase diagram.

\newpage
\section{Harmonic Raman spectra}

The following harmonic Raman spectra analysis was performed using the phonopy-spectroscopy code.\cite{skelton2017lattice} Peak positions were evaluated assuming harmonic vibrational behaviour and using the PBEsol functional. Dielectric tensors were evaluated using the PBE functional. A light basis set with a minimum density of 5 k-grids per \AA$^{-1}$ was used throughout. A Lorentzian of width 0.05cm$^{-1}$ is used to broaden the peaks. 


\ce{BaZrS3} in the the \hmn{Pnma} phase has 60 optic phonon modes, of which 24 are Raman active: $\Gamma$ = 7A$_g$ $\bigoplus$ 5B$_{1g}$ $\bigoplus$ 7B$_{2g}$ $\bigoplus$ 5B$_{3g}$. \\
\begin{figure}[H]
    \centering
    \includegraphics[width=0.5\linewidth]{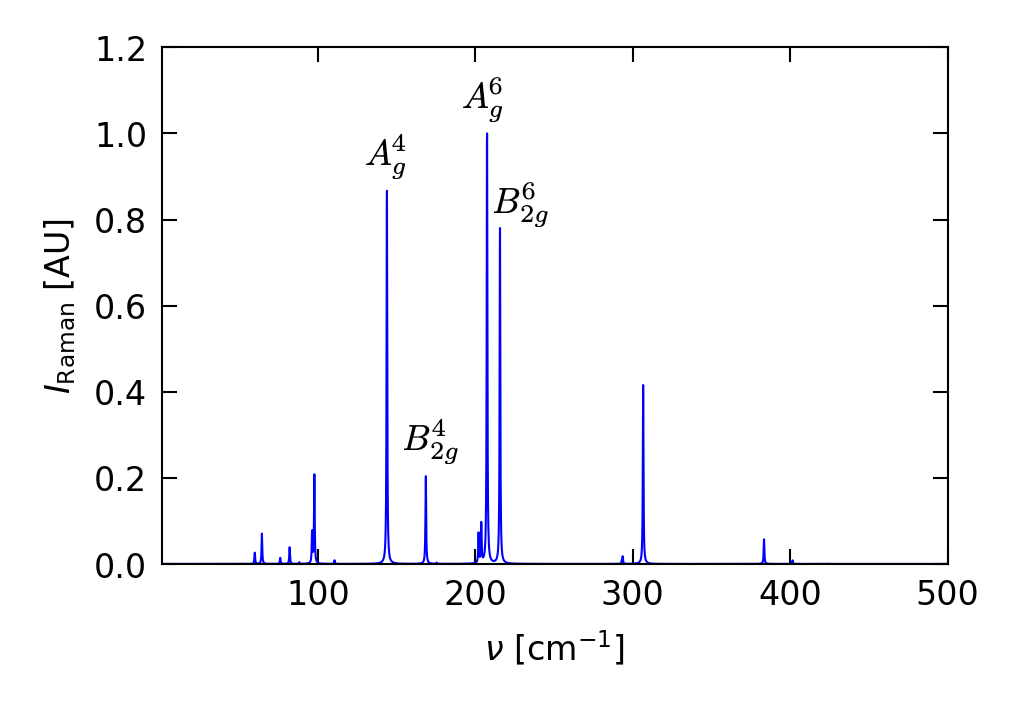}
    \caption{Harmonic first-order Raman spectra of the $Pnma$ phase}
    \label{sfig:Raman_Pnma}
\end{figure}
\begin{table}[H]
    \centering
    \begin{tabular}{ccc}
     Frequency [cm$^{-1}$]   & Intensity [\AA$^{4}$ amu$^{-1}$] & Mode Symmetry\\
57.38246  &  31.294167  &  A$_{g}$  \\
62.740697 &  74.397711  &  B$_{2g}$  \\
72.865691 &  20.712113  &  A$_{g}$  \\
74.878717 &  0.428877   &  B$_{3g}$  \\
79.451252 &  35.213175  &  B$_{2g}$  \\
82.730798 &  4.047662   &  B$_{1g}$  \\
89.396457 &  81.906317  &  B$_{2g}$  \\
94.004782 &  202.648574 &  A$_{g}$  \\
105.117071&  8.151943   &  B$_{1g}$  \\
139.776095&  927.058612 &  A$_{g}$  \\
157.205497&  0.129784   &  B$_{3g}$  \\
162.91409 &  192.607928 &  B$_{2g}$  \\
166.327512&  3.336429   &  A$_{g}$  \\
171.34857 &  2.338391   &  B$_{1g}$  \\
198.430549&  98.18659   &  B$_{2g}$  \\
200.803735&  93.792521  &  B$_{3g}$  \\
204.900935&  1063.364972&  A$_{g}$  \\
213.109461&  766.788035 &  B$_{2g}$  \\
289.681849&  19.314493  &  B$_{1g}$  \\
289.943548&  8.201392   &  B$_{3g}$  \\
304.087743&  361.586611 &  A$_{g}$  \\
389.08038 &  48.066782  &  B$_{1g}$  \\
400.954212&  7.889141   &  B$_{2g}$  \\
419.617256&  0.634279   &  B$_{3g}$  \\
    \end{tabular}
    \caption{Raman frequencies and mode symmetries for the \hmn{Pnma} phase}
    \label{tab:Raman_Pnma}
\end{table}
\ce{BaZrS3} in the \hmn{I4/mcm} phase has 30 optic phonon modes, of which 7 are Raman active: $\Gamma$ =  3E$_g$ $\bigoplus$ A$_{1g}$ $\bigoplus$ 2A$_{2g}$ $\bigoplus$ B$_{1g}$ $\bigoplus$ B$_{2g}$. \\ 
\begin{figure}[H]
    \centering
    \includegraphics[width=0.5\linewidth]{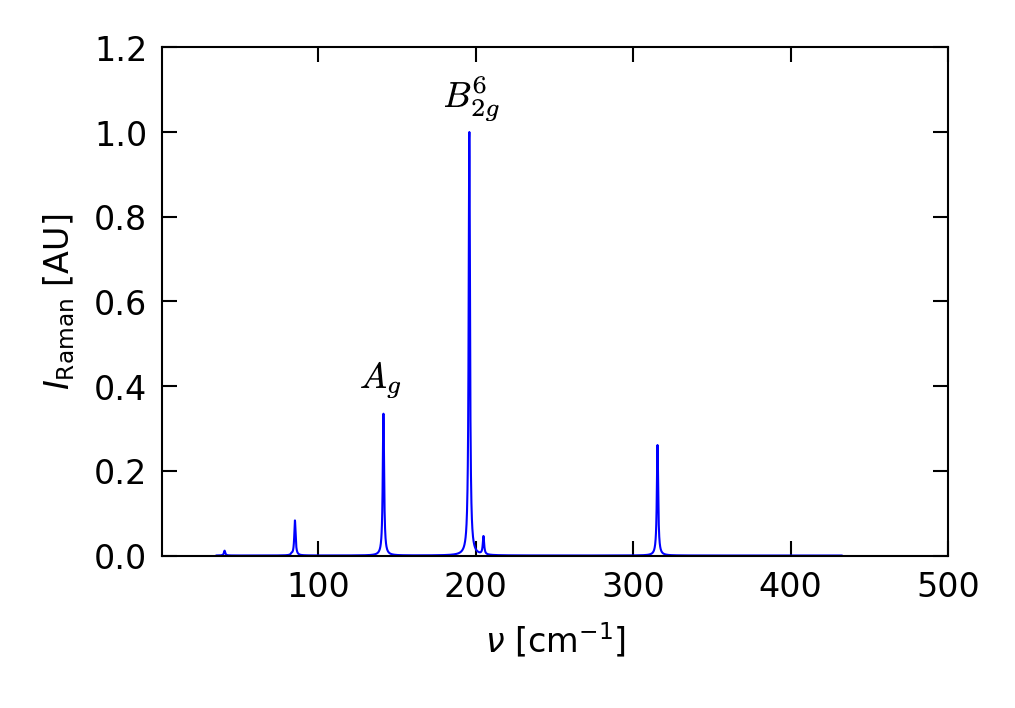}
    \caption{Harmonic first-order Raman spectra of the $I4/mcm$ phase}
    \label{sfig:Raman_I4_mcm}
\end{figure}
\begin{table}[H]
    \centering
    \begin{tabular}{ccc}
 Frequency [cm$^{-1}$]   & Intensity [\AA$^{4}$ amu$^{-1}$] & Mode Symmetry\\
    40.762563          &          3.683262 &    E$_g$    \\
    40.762563          &          3.687507 &    E$_g$    \\
    42.565308          &          0.000000 &    A$_{2u}$    \\
    69.245268          &          0.000000 &    E$_u$     \\
    69.245268          &          0.000000 &    E$_u$     \\
    83.275265          &          0.896384 &    E$_g$     \\
    83.275265          &          0.894630 &    E$_g$     \\
    85.401421          &         53.217047 &    B$_{2g}$    \\
   100.908270          &          0.000000 &    E$_u$     \\
   100.908270          &          0.000000 &    E$_u$     \\
   126.898251          &          0.000000 &    E$_u$     \\
   126.898251          &          0.000000 &    E$_u$     \\
   137.013961          &          0.000000 &    A$_{2u}$    \\
   141.559757          &        216.862006 &    A$_{1g}$    \\
   155.032805          &          0.000000 &    B$_{1u}$    \\
   196.069483          &        645.987936 &    B$_{2g}$    \\
   205.046737          &         13.904421 &    E$_g$     \\
   205.046737          &         13.919185 &    E$_g$     \\
   207.325511          &          0.000000 &    A$_{1u}$    \\
   230.492719          &          0.000000 &    E$_u$     \\
   230.492719          &          0.000000 &    E$_u$     \\
   281.507146          &          0.000044 &    A$_{2g}$    \\
   302.441159          &          0.000000 &    A$_{2u}$    \\
   303.618329          &          0.000000 &    E$_u$     \\
   303.618329          &          0.000000 &    E$_u$     \\
   315.495626          &        167.934156 &    B$_{1g}$    \\
   427.229088          &          0.000392 &    A$_{2g}$    \\
    \end{tabular}
    \caption{Raman frequencies and mode symmetries for the \hmn{I4/mcm} phase}
    \label{tab:Raman_I4_mcm}
\end{table}
\ce{BaZrS3} in the \hmn{Pm-3m} phase has 15 optic phonon modes however, due to Raman selection rules, none of the modes are first-order Raman active. Any Raman signal observed from this phase derive from second-order Raman effects. 

\newpage
\section{Static structure factor}

\begin{figure}[H]
    \centering
    \includegraphics[width=0.8\linewidth]{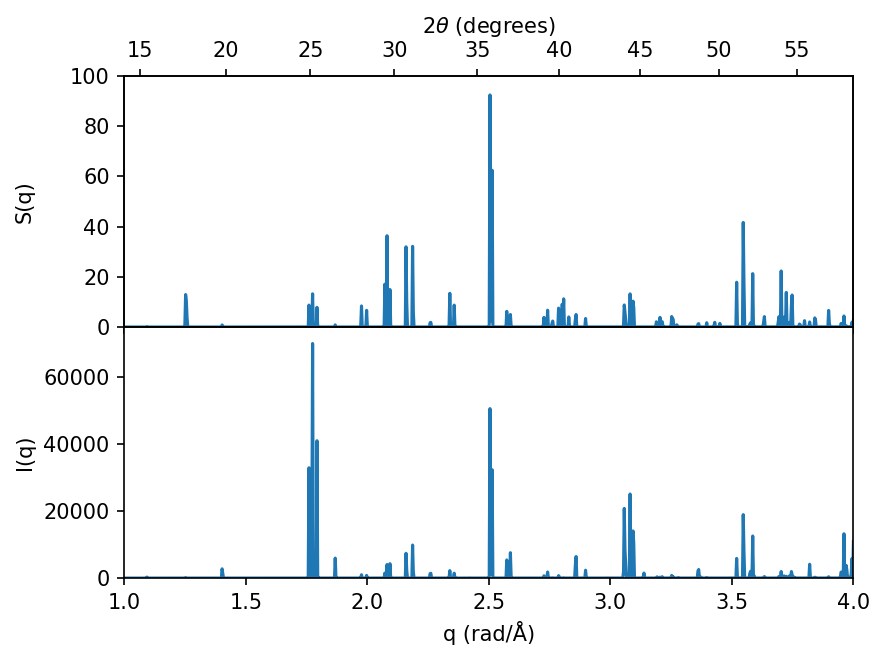}
    \caption{Comparison of the static structure factor pattern (S(q)) to the x-ray diffraction pattern (I(q)). }
    \label{fig:S_I_compare}
\end{figure}

X-ray scattering intensities are equivalent to static structure factors weighted with species- and source- dependent form factors. We demonstrate the effect of including the form factors through direct comparison in \autoref{fig:S_I_compare}. A number of peaks are not present in the scattering intensity, most noticeably at a scattering angle of \ang{17.5}.

\begin{figure}[H]
    \centering
    \includegraphics[width=0.6\linewidth]{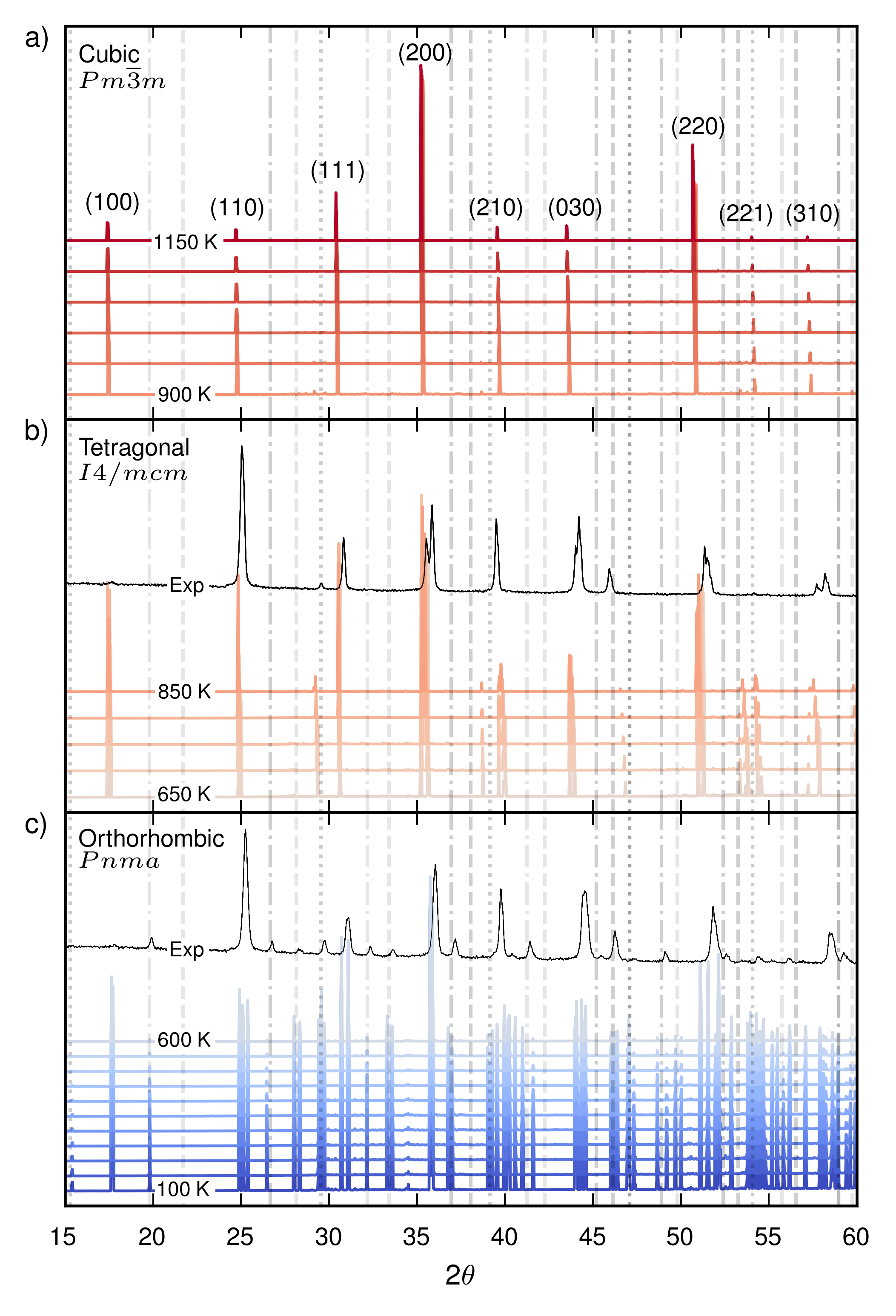}
    \caption{Static structure factor, $S(q)$, evaluated for three 
    \ce{BaZrS3} polymorphs from \qtyrange{100}{1150}{\kelvin} in intervals of \qty{50}{\kelvin}. A Cu K$\alpha$ value of \qty{1.5406}{\angstrom} was used for the $q$ to $\theta$ conversion.
    Cubic \hmn{Pm-3m} peaks are indexed. 
    Superlattice peaks (up to the fifth Brillouin Zone) associated with R-, M- and X-point distortions are indexed and marked with dotted, dashed, and dash-dotted lines, respectively. 
    Experimental XRD data of the orthorhombic (\qty{303}{\kelvin}) and tetragonal (\qty{923}{\kelvin}) phases from Ref.~\citenum{bystricky2024thermal} are plotted in black.}
    \label{fig:XRD_full}
\end{figure}

\newpage
The temperature-dependent static structure factors are displayed in \autoref{fig:XRD_full}. In contrast to the main text, superlattice peaks up to the fifth Brillouin Zone are displayed.

\newpage

\section{Partial static structure factors and X-ray diffraction scattering intensities}  

\begin{figure}[H]
    \centering
    \includegraphics[width=0.8\linewidth]{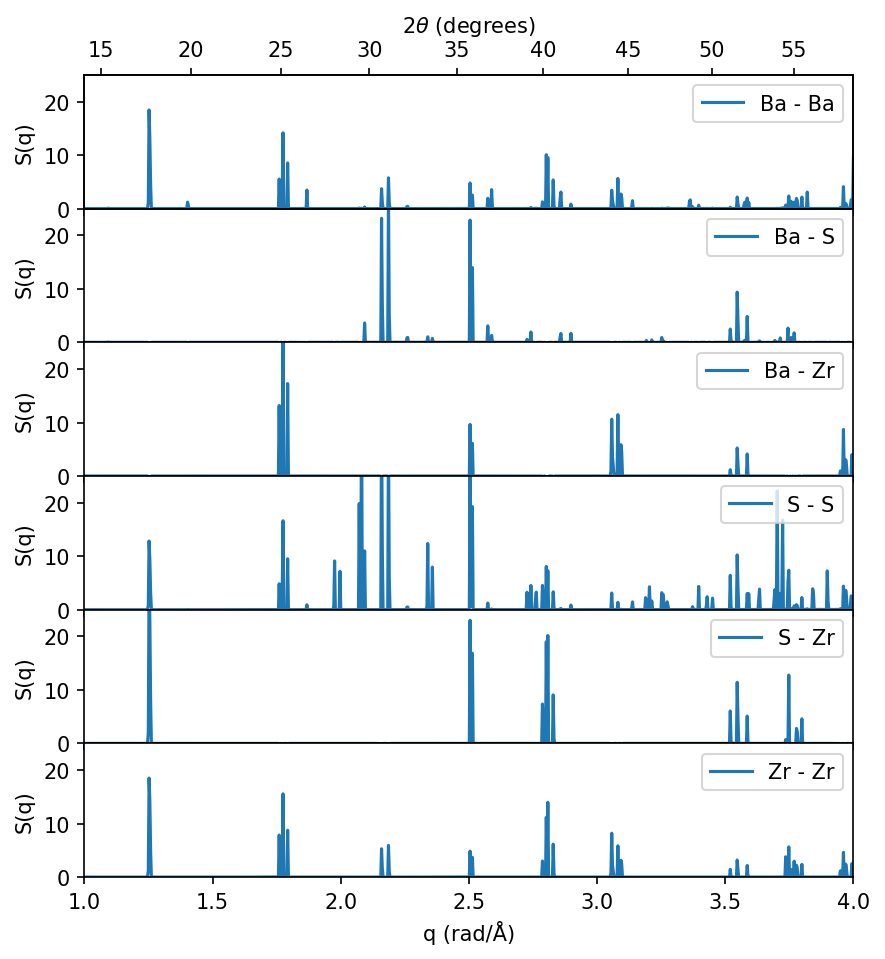}
    \caption{Partial static structure factors, $S(q)$, for \ce{BaZrS3} in the \hmn{Pnma} phase at \qty{300}{\kelvin} and \qty{0}{\pascal}.}
    \label{fig:atomic-weighted-structure}
\end{figure}

\begin{figure}[H]
    \centering
    \includegraphics[width=0.8\linewidth]{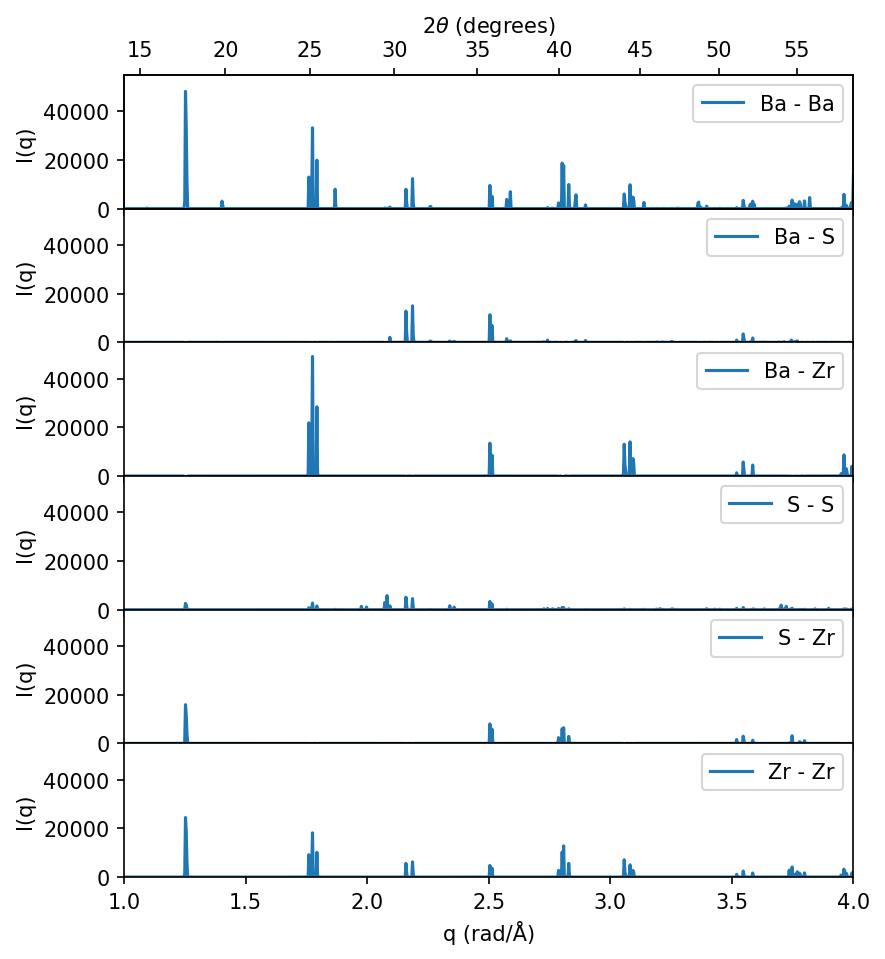}
    \caption{Partial scattering intensities, $I(q)$, for \ce{BaZrS3} in the \hmn{Pnma} phase at \qty{300}{\kelvin} and \qty{0}{\pascal}.}
    \label{fig:atomic-weighted-XRD}
\end{figure}

We can decompose the static structure factors for the various atom types in the system to compute partial static structure factors. In \autoref{fig:atomic-weighted-structure} we plot the partial structure factor of the \hmn{Pnma} phase at \qty{300}{\kelvin} and \qty{0}{\pascal}. Similarly, we can decompose the X-ray diffraction scattering intensity $I(q)$. This is displayed in \autoref{fig:atomic-weighted-XRD}.

In \autoref{fig:atomic-weighted-structure} and \autoref{fig:atomic-weighted-XRD} there is a peak at $q = 1.4$ for the Ba-Ba partials. This corresponds to a distortion at the X-point which involves atomic displacement of the Ba species only.
 
\clearpage
\phantomsection
\addcontentsline{toc}{section}{\listreferencename}

\bibliography{references}